# Malicious Mode Attack on EV Coordinated Charging Load and MIADRC Defense Strategy

Yichen Zhou [a, c], Weidong Liu [b, a], Jing Ma[c, d*], Xinghao Zhen[a, c], and Yonggang Li[a, c]

[a] *Electrical and Electronic Engineering Department, North China Electric Power University, No. 619 Yonghua North Road, Baoding, 071003, China*
[b] *State Grid Tianjin Electric Power Company, Tianjin, 300000, China*
[c] *State Key Laboratory of Alternate Electrical Power System with Renewable Energy Sources, North China Electric Power University, No. 2 Beinong Road, Beijing, 102206, China*
[d] *China Institute of Energy and Transportation Integrated Development, North China Electric Power University, No. 2 Beinong Road, Beijing, 102206, China*

***Abstract*—The Internet of Things (IoT) provides a salient communication environment to facilitate the coordinated charging of electric vehicle (EV) load. However, as IoT is connected with the public network, the coordinated charging system is in a low-level cyber security and greatly vulnerable to malicious attacks. This paper investigates the malicious mode attack (MMA), which is a new cyber-attack pattern that simultaneously attacks massive EV charging piles to generate continuous sinusoidal power disturbance with the same frequency as the poorly-damped wide-area electromechanical mode. Thereby, high amplitude forced oscillations could be stimulated by MMA, which seriously threats the power system stability. First, the potential threat of MMA is clarified by investigating the vulnerability of the IoT-based coordinated charging load control system, and an MMA process like Mirai is pointed out as an example. And then, based on the attack process, an MMA model is established for impact analysis where expressions of the mean and stochastic responses of the MMA forced oscillation are derived to discover main impact factors. Further, to mitigate the impact of MMA, a defense strategy based on multi-index information active disturbance rejection control is proposed to improve the stability and anti-disturbance ability of the power system, which considers the impact factors of both mode damping and disturbance compensation. Simulations are conducted to verify the existence and characteristics of MMA threats, and the efficiency of the proposed defense strategy is also validated.***

## I. Introduction

The increasing charging demand of the electric vehicle (EV) has put tremendous pressures on the power supply. Accordingly, coordinated charging strategies based on Internet of things (IoT) technology is developing rapidly, e.g., the smart charging (V1G) and the vehicle-to-grid (V2G), which efficiently exploits the flexibility and controllability of the EV.

The charging network fitted out with the Internet connection and the satellite navigation can coordinately control the EV load. Moreover, the transportation network is connected closer to the power grid with the widespread of the IoT in EV, charging network, smart grid, etc. Now the EV can obtain more flexible, convenient, and friendly electric power through the intelligent integrated controller of industrial internet of things (IIoT). Thus, the coordinated charging network serves as a key part of the smart grid, forming a complex large system fused with power, information and the communication network.

In the sophisticated information-sharing IIoT, the security issue poses various challenges to the safe operation of the power infrastructure. In 2010, Iran's industrial infrastructure is hit by Stuxnet worm via the supervisory control and data acquisition (SCADA) system [1]. In 2015, hours of Ukraine blackouts were caused by cyber-attacks on three region's distribution network company [2]. The IOT virus has broken out since 2016, e.g., Mirai, Haijime, Persirai, BrickerBot, Okiru, et cetera. They exploit vulnerabilities and produce multiple cyber-attack ways such as distributed denial of service (DDoS) attack, malicious data [4], malicious software [5], and password intrusion [6]. The common feature is that malicious attackers control industrial equipment to break the power energy balance through cyber-attacks, causing the destruction of power system stability. Therefore, to reduce damage, researchers should race against attackers to find potential cyber-attack vulnerabilities and provide defense strategies in advance.

A large-scale cyber-attack directly against power-consuming devices, compared with the traditional cyber-attack against transmission and distribution devices, is more critical and harder to defend due to the diversity and dispersion of power-consuming. Recently, EVs are aggregated to participate in peak cut [7], frequency regulation [8] and power system dynamic response [9] through coordinated charging networks. However, the aggregated EVs are also at risk of malicious control. [10] summarized typical cyber-attacks via grid-tied converters. Moreover, EVs can also be used as a bridge or platform to launch malicious programs [11]. Thus, it is necessary to protect the energy-Internet-based V2G communication from cyber-attacks [12]. However, it is difficult for the cyber system to propose a unified secure mechanism that could defend against all cyber-attacks. Hence, besides improving cyber security, the physical power grid should also enhance its survivability under cyber-attacks.

Motivated by above issues, this paper investigates malicious mode attack (MMA), a new cyber-attack pattern, which utilizes the coordinated charging load control system (CCLCS) as the tool to send the malicious code of the forced oscillation control command. It can create a high-amplitude forced oscillation according to the weakly damped wide-area electromechanical mode, thereby destroying the stability of the system. Thus, the MMA model is established, followed by analysis on the characteristics of MMA. Moreover, it is difficult to actively eliminate this risk through the existing cyber technology of defense or the physical technology of forced oscillation positioning. Therefore, this paper provides a multi-information active distribution rejection control (MIADRC) method for the physical smart grid to enhance its robustness, thereby passively suppressing the risk of forced oscillation.

Main contributions of this paper are (1) the MMA threat through the coordinated charging is discovered and investigated, (2) the model of MMA via the coordinated charging is established and characteristics are concluded through theoretically analysis, (3) and an MIADRC strategy is provided to enhance the defense capability against MMA.

This paper is organized as follows. Section II describes the CCLCS structure and reveals the MMA threat. Section III introduces the model of MMA and analyzes MMA characteristics. Section IV provides the MIADRC defense strategy. Section V validates the MMA threat, its characteristics, and the effectiveness of the proposed defense strategy, followed by conclusions given in Section VI.

## II. Coordinated Charging Load Control Systems and Malicious Mode Attacks

The coordinated charging guides the charging behavior of EVs to follow the economic dispatch command via CCLCS [13]. Thus, this section will provide a bird's eye view of CCLCS and investigate its vulnerability to the MMA threat, followed by the definition, attack process and malicious command of MMA to further demonstrate the MMA threat.

### *A. Coordinated Charging Load Control System*

CCLCS consists of three parts: power conversion, communication network, and energy management. By making the EV charging load act as the demand response resource, CCLCS downward controls the charging load of EVs, and upward participates in the coordinated schedule of "source-grid-load". A typical CCLCS is shown in Fig. 1.

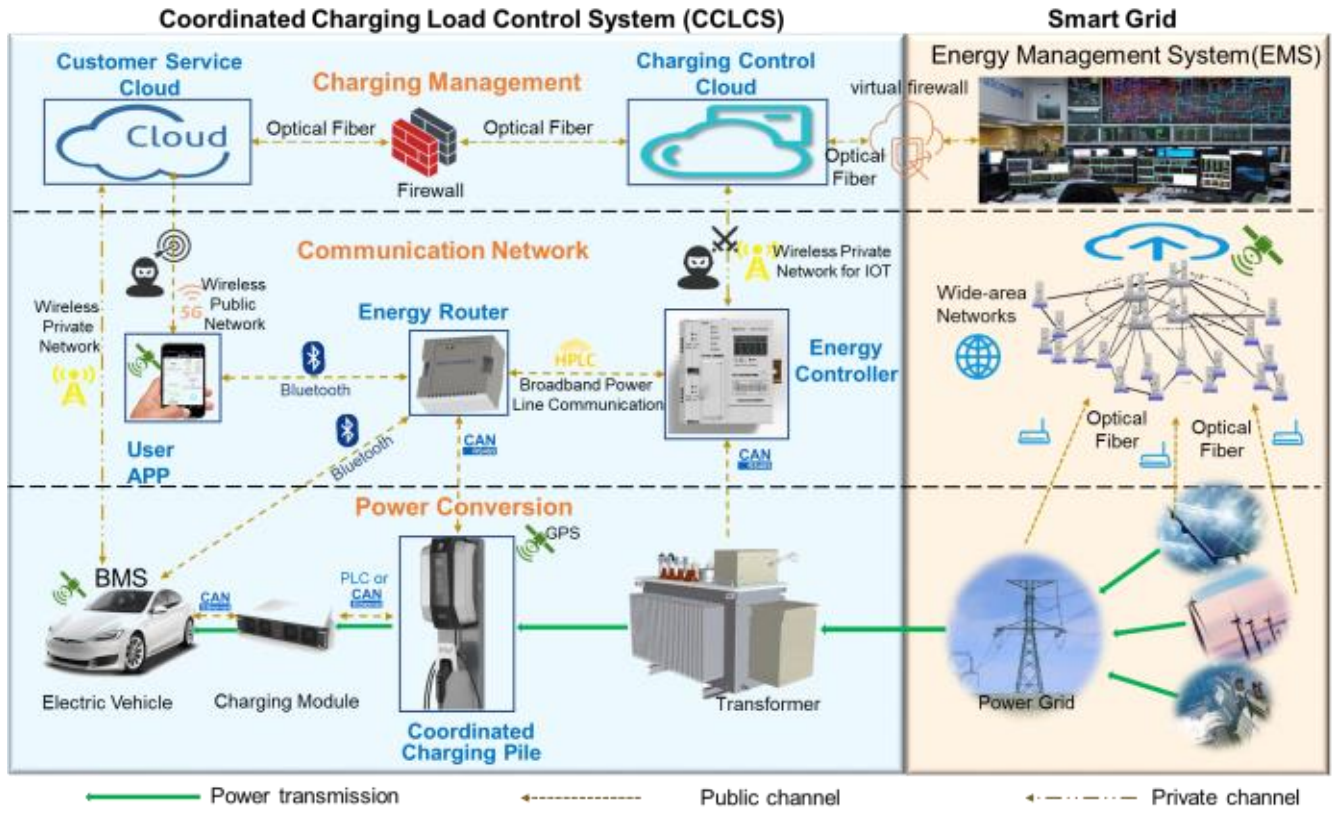

Fig. 1. A typical CCLCS.

As is shown in Fig. 1, main devices of CCLCS include coordinated charging pile, user applications (Apps), energy router, energy controller, charging control cloud, and customer service cloud. Main systems contained in CCLCS are energy management system (EMS), distribution automation system, global position system (GPS), IoT, battery management system (BMS). By these devices and systems, the controlled electrical power is transmitted to the charging pile via generator, transmission line, distribution network, and transformer under the joint control of CCLCS and EMS.

For the control process of CCLCS, the charging control cloud develops a coordinated charging power control strategy that considers both the dispatch commands from the dispatch control center and the customer demands from the customer service cloud, and passes it to the energy controller. Then the energy controller optimizes the control strategy based on the distribution network information and the power constraints of transformers, and sends it to the energy router. Next, the energy router combines the received control command with the state-of-charge (SOC) given by BMS to optimize the final order. This order will finally be executed by the charging pile in the power exchanging process with the EVs.

To facilitate the above control process, first, the information and communication technology (ICT) technique is adopted by CCLCS to send and received information, including Bluetooth, CAN, PLC, satellite communication, mobile communication, etc. Then, with the contribution of ICT, components of the IoT operating system (OS), including energy controller, energy router and coordinated charging piles, can be connected for the implementation of communication and control functions, where main communication modes are illustrated in Table I. Besides, owing to GPS or NTP, all devices in Fig. 1 can achieve time synchronization in a high accuracy level.

TABLE I

Communication Modes

| **Component 1** | **Component 2** | **Communication Modes** |
|---|---|---|
| Charging control cloud | Energy controller | Wireless private network for IoT |
| Charging control cloud | EMS | Optical fiber with virtual firewall |
| Energy controller | Energy router | HPLC and (or) NB-IoT |
| Energy router | Charging pile | CAN |
| Charging pile | EV | PWM/PLC/CAN |
| EV, user App | Energy router | Bluetooth |
| EV, user App | Customer service cloud | Wireless network |
| Customer service cloud | Charging control cloud | Optical fiber with firewall |

However, a huge MMA threat is buried. Traditionally, the power system adopts the private optical fiber as the communicate channel to keep a high-level communication security, which is strictly isolated from the public communication. However, as shown in Table I, CCLCS exchanges data with public consumers through the public communication network, CCLCS is exposed to cyber attackers in the public communication network. For instance, since connected to user apps from public networks, the energy router and customer service cloud are at high risks of being attacked. Unfortunately, it is very difficult for the power supplier to ensure the security of all public network information. Finally, as terminals of CCLCS, massive charging piles are easy to be manipulated to launch MMA by cyber-attackers, seriously threating power system stability.

### *B. Malicious Mode Attacks*

In this paper, MMA is defined as a cyber-attack that sends Trojan horse viruses containing a malicious load control command which controls the behavior of large-scale EV charging loads to simultaneously act as a continuous sinusoidal power disturbance with the same frequency as the weakly damped inter-area electromechanical oscillation mode.

Once the vulnerability in public network is found, the attack process of MMA is not a big problem for hackers. A junior attacker can penetrate the customer service cloud through the user APP, and then invade the charging control cloud system to send Trojans through the firewall vulnerabilities. Moreover, professional hackers may hack the charging control system directly to send Trojans or the malicious coordinated charging control commands.

The MMA attacker can first penetrate into CCLCS, then infect devices in CCLCS, and finally attack charging piles simultaneously. To show this attack process clearly, a typical MMA like Mirai is taken as an example and illustrated in Fig. 2. And to save space, attack steps of Fig. 2 are summarized and given in Algorithm 1. The penetration, infection, and attack steps of MMA correspond to blue, red, and yellow lines in Fig.2, respectively, and also correspond to the first, second, and third parts of Algorithm 1, respectively.

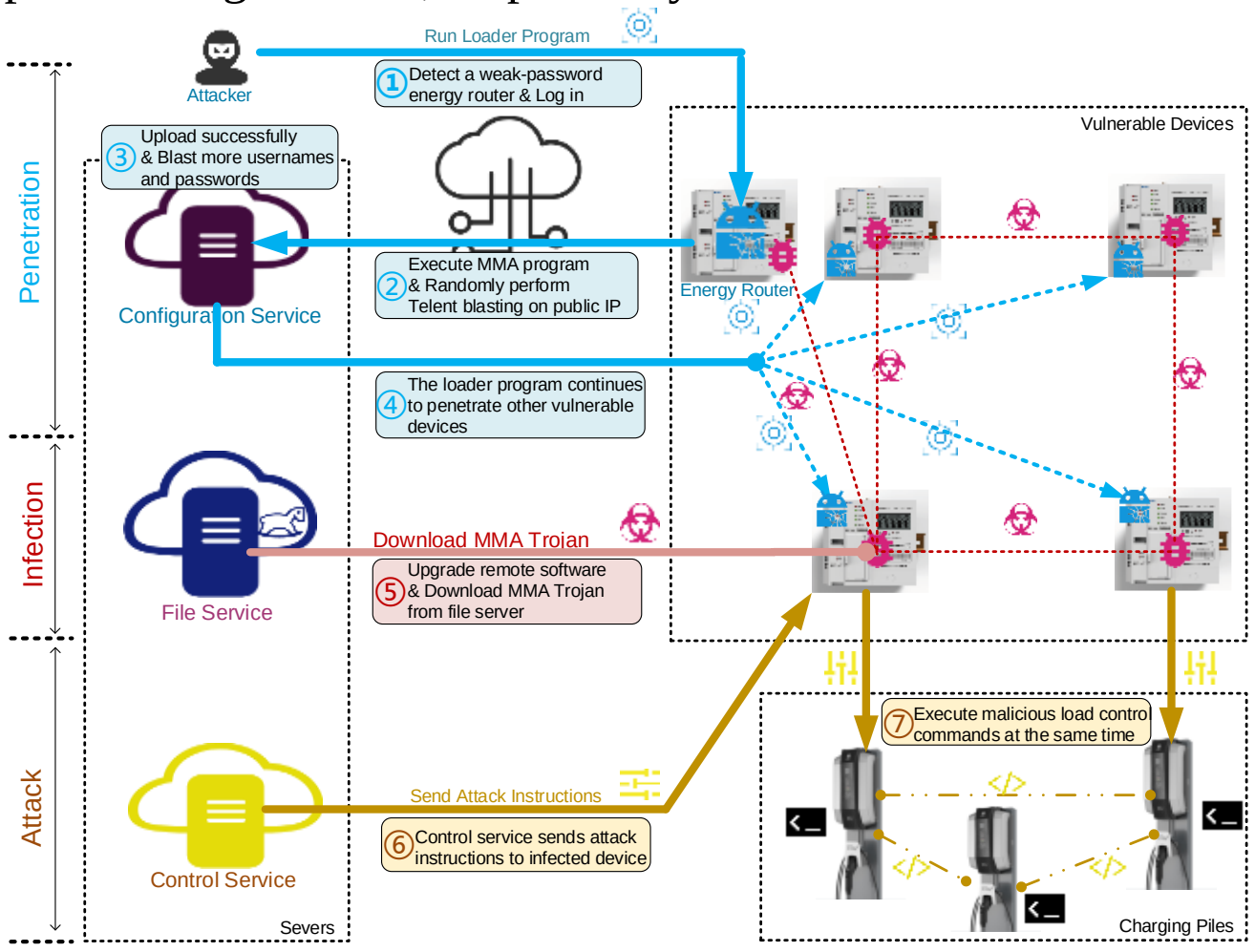

Fig. 2. A typical process of MMA like Mirai.

**Algorithm 1**: Process of MMA like Mirai.

| | | |
|---|---|---|
| **Penetration** | 1 | The attacker runs the loader program to detect a weak-password energy router, and then logs in. |
| | 2 | The penetrated energy router executes the MMA program and randomly performs Telnet blasting on the public IP. |
| | 3 | More usernames and passwords of devices are blasted through the configuration sever. |
| | 4 | The loader program continues to penetrate other vulnerable devices. |
| **Infection** | 5 | The infected devices download the MMA Trojan program from the file server after upgrading the remote software. |
| **Attack** | 6 | The attacker obtains the weakly-damped mode from FNET and send the malicious load control command to the control server. And the control server sends attack instruments to infected devices. |
| | 7 | The malicious load control commands are triggered by the MMA Trojans and executed simultaneously at a unified GPS clock via the charging piles. |

Through the above attack process, CCLCS commands can be maliciously manipulated by MMA, which means the normal power control command of the $i^{\text{th}}$ charging pile $P_{c0i}$ can be intercepted and tampered with. A continuous power disturbance command is generated by MMA Trojans and added to $P_{c0i}$ so that the final power control command $P_{ci}$ is

$$P_{ci}(t) = P_{c0i}(1 + I\%\cos\omega t) \tag{1}$$

where *I%* is the ratio of the cosine disturbance peak value to the normal $P_{c0i}$, and $\omega$ is the frequency of this disturbance which can be identified by FNET or μPMU from an ordinary household electrical outlet. The above parameters of the malicious load control command can be sent to the Trojan in 5 minutes before the attack, and then numerous EVs will launch the attack simultaneously using the unified GPS clock.

Finally, under the malicious manipulation of MMA, power fluctuations of numerous EV loads will cause a high-amplitude forced oscillation. Hence, the MMA threat to power system stability is formed.

## III. MMA Model for Characteristic Analysis

Conducting theoretical analysis to determine key factors is the basis for eliminating the MMA threat. In this section, a mathematical model is first built to represent the power system under MMA. And then, theoretical analysis is conducted to investigate characteristics of stimulated forced oscillation to MMA and propose key points of MMA defense.

### *A. Mathematical Model for Power System under MMA*

The charging pile converter includes a VIENNA rectifier for ac-dc conversion and a full bridge LLC series resonant converter for the dc-dc conversion [14]. As this paper focuses on dynamics of the electromechanical time scale, an electromechanical dynamic model of a normal charging pile is built (Fig. 3) which includes phase locked loop (PLL), power control, simplified current control [15], and Park transformation with fast dynamics of the DC circuit ignored. It is expressed as

$$\begin{cases} \dot{x}_{PLL} = -u_{gd} \\ \dot{\theta}_{PLL} = -K_{p3}u_{gd} + K_{i3}x_{PLL} \\ \dot{i}_q = (i_{qref} - i_q)/\tau_1 \end{cases} \text{ and } \begin{cases} \dot{x}_1 = K_{i1}(P_{ref} - P_e) \\ \dot{i}_d = \dfrac{1}{\tau_2}(i_{dref} - i_d) \\ \dot{x}_2 = K_{i2}(Q_{ref} - Q_e) \end{cases} \tag{2}$$

where $x_{PLL}$ and $\theta_{PLL}$ are state variables of PLL, $i_d$ and $i_q$ are $d$-axis and $q$-axis currents injected into the transmission grid, $i_{qref} = x_1 + K_{p1}(P_{ref} - P_3)$, $i_{dref} = x_2 + K_{p2}(Q_{ref} - Q_e)$, $P_{ref}$ and $Q_{ref}$ are commands for active and reactive power control, three PI controls are adopted for PLL and the power control, and $\tau_1$ and $\tau_2$ are the time delay introduced by the inner current control loop.

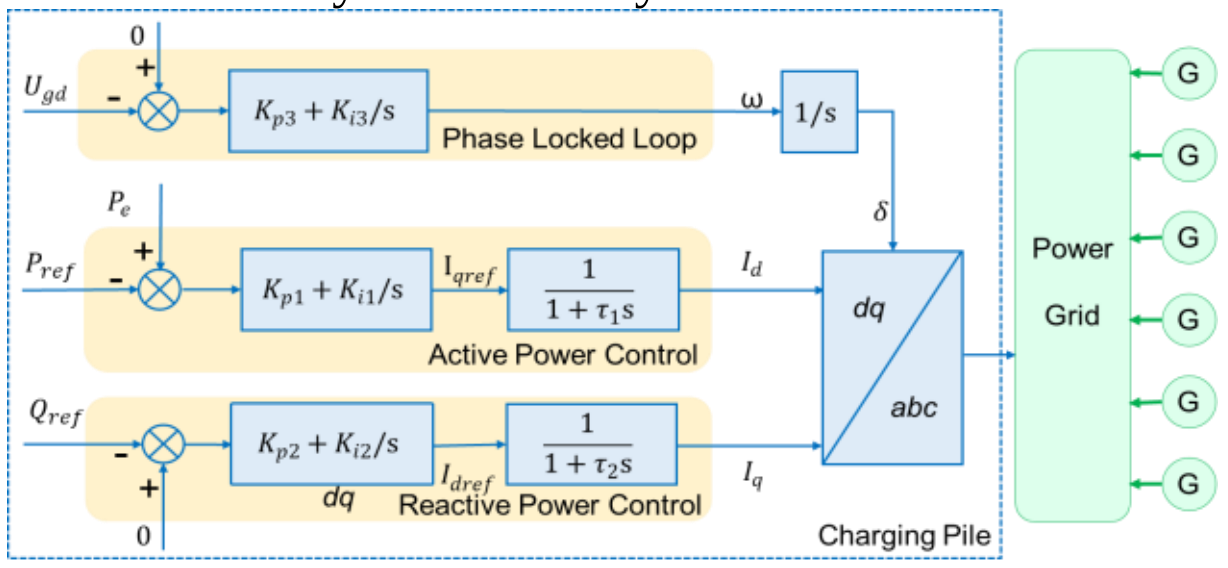

Fig. 3. A simple electromechanical dynamic model of the charging pile.

During normal operation, the active power command $P_{ref}$ of the $i^{\text{th}}$ charging pile is $P_{c0i}$. In the large target scope of MMA with $n$ charging piles, we will obtain the aggregated active power command $P_{a0}$ by summing $P_{c0i}$, $i$=1,⋯,$n$.

Under MMA, according to Section II. B, the active power commands of charging piles are manipulated by MMA. As given in (1), $P_{c0i}$ is tampered with to $P_{ci}$. Thus, $P_{a0}$ will become $P_a$ which can be computed by

$$P_a(t)=\sum_{i=1}^{n}P_{c0i}(t)(1+I\%\cos\omega t)=P_{a0}(t)(1+I\%\cos\omega t) \quad (3)$$

where $I\%$ and $\omega$ are constant and determined before MMA as explained below (1), and $P_{a0}(t)=\sum_{i=1}^{n}P_{c0i}(t)$ . Because the charging demand and charging behavior of each EV are stochastic, $P_{c0i}(t)$ and the aggregated value $P_{a0}(t)$ are stochastic. However, the forced oscillation stimulated by MMA only lasts tens of seconds, and $P_{a0}(t)$ remains almost constant during this period. Hence, $P_{a0}(t)$ can be considered as a stationary stochastic process $P_{a0}$ , and $P_{a0}$ approximately obeys normal distribution [16].

Thus, considering MMA, (2) can model both single and aggregated massive charging piles by assigning different active commands. Once, MMA is launched, for a single charging pile, $P_{ref}=P_{ci}(t)$, and for aggregated charging piles, $P_{ref}=P_a(t)$.

Then, the aggregated charging pile model is added into the power system model to build the mathematical model of the whole power system for further theoretical analysis on MMA, which can be expressed as

$$\begin{cases}\dot{\boldsymbol{x}}(t)=\boldsymbol{A}\boldsymbol{x}(t)+\boldsymbol{B}\boldsymbol{u}(t)\\ \boldsymbol{y}(t)=\boldsymbol{C}\boldsymbol{x}(t)+\boldsymbol{D}\boldsymbol{u}(t)\end{cases} \quad (4)$$

where, $\boldsymbol{x}\in \mathrm{R}^n$ and $\boldsymbol{y}\in \mathrm{R}^m$ are state and output variable vectors of the whole system, $\boldsymbol{u}(t)=I\%P_{a0}\cos(\omega t+\varphi)$ is the aggregated malicious command from MMA, and $\boldsymbol{A}\in \mathrm{R}^{n\times n}$ , $\boldsymbol{B}\in \mathrm{R}^{n\times l}$ , $\boldsymbol{C}\in \mathrm{R}^{m\times n}$ , and $\boldsymbol{D}\in \mathrm{R}^{n\times l}$ are system matrices.

### *B. Analysis on Forced Oscillation Stimulated by MMA*

Due to $P_{a0}$ , the disturbance $\boldsymbol{u}(t)$ injected by MMA is a stochastic process. Thus, a stochastic process analysis is carried out to determine the stochastic characteristics of the stimulated forced oscillation by MMA.

For convenience, (4) is changed from coordinates $\boldsymbol{x}$ to the main coordinates $\boldsymbol{z}$ by using the modal superposition method

$$\boldsymbol{x}=\boldsymbol{U}\boldsymbol{z} \quad (5)$$

where, $\boldsymbol{U}\in \mathrm{R}^{n\times n}$ is the right eigenvector of the state matrix $\boldsymbol{A}$ .

As $\boldsymbol{A}$ is usually semisimple for the power system, the modal transformation system is completely decoupled. In the new coordinates $\boldsymbol{z}$ , the transformed state equation is expressed as

$$\begin{cases}\dot{\boldsymbol{Z}}(t)=\boldsymbol{\Lambda}\boldsymbol{Z}(t)+\boldsymbol{\Psi}\boldsymbol{u}(t)\\ \boldsymbol{Y}(t)=\boldsymbol{\Phi}\boldsymbol{Z}(t)+\boldsymbol{D}\boldsymbol{u}(t)\end{cases} \quad (6)$$

where, $\boldsymbol{\Lambda}=\boldsymbol{U}^{-1}\boldsymbol{A}\boldsymbol{U}=\boldsymbol{V}^T\boldsymbol{A}\boldsymbol{U}$ , $\boldsymbol{\Psi}=\boldsymbol{U}^{-1}\boldsymbol{B}=\boldsymbol{V}^T\boldsymbol{B}$ , and $\boldsymbol{\Phi}=\boldsymbol{C}\boldsymbol{U}$ .

For the stochastic characteristics of the simulated forced oscilaltion, by using the complex modal superposition method, it can deduced that the mean response of $\boldsymbol{z}(t)$ to $\boldsymbol{u}(t)$ is

$$\begin{aligned}\mathrm{E}(\boldsymbol{z}(t))&=e^{\Lambda t}\boldsymbol{z}(0)+\int_0^t e^{\Lambda(t-\tau)}\boldsymbol{\Psi}\bar{P}_{a0}\left(\frac{e^{j\omega\tau+j\varphi}+e^{-j\omega\tau-j\varphi}}{2}\right)d\tau+\boldsymbol{D}\boldsymbol{u}(t)\\ &=e^{\Lambda t}\boldsymbol{z}(0)+\frac{\bar{P}_{a0}}{2}[e^{j\varphi}(\boldsymbol{\Lambda}-j\omega\boldsymbol{I})^{-1}(e^{\Lambda t}-e^{j\omega\boldsymbol{I}t})+\\ &\quad e^{-j\varphi}(\boldsymbol{\Lambda}+j\omega\boldsymbol{I})^{-1}(e^{\Lambda t}-e^{-j\omega\boldsymbol{I}t})]\boldsymbol{\Psi}+\boldsymbol{D}\boldsymbol{u}(t)\end{aligned} \quad (7)$$

where $\bar{P}_{a0}=I\%E(P_{a0})$ is the mean of $I\%P_{a0}$ .

In (7), the first item is a free oscillation that can be ignored when the initial state is zero, and the third item can also be omitted since $D$ is zero. Therefore, the mean response of $y_k$ (the $k^{\text{th}}$ element of $\mathbf{y}(t)$) to $\boldsymbol{u}(t)$ is in a simple form expressed as

$$E(y_k(t))=\sum_{i=1}^{n-1}2\bar{P}_{a0}A_{ki}e^{\alpha_i t}\cos(\omega_i t+\alpha_{ki})-2\bar{P}_{a0}B_k\cos(\omega t+\beta_k) \quad (8)$$

where $A_{ki}=|\phi_{ki}||\psi_i|\sqrt{\dfrac{(\alpha_i\cos\tilde{\varphi}-\omega\sin\tilde{\varphi})^2+\omega_i^2\cos^2\tilde{\varphi}}{\left(\alpha_i^2+(\omega_i^2-\omega^2)\right)^2+4\alpha_i^2\omega^2}}$ , $B_k=\sqrt{\sum_{i=1}^{n}B_{ki}^{\ 2}}$ , $B_{ki}=|\phi_{ki}||\psi_i|\sqrt{\dfrac{(\alpha_i\cos\tilde{\varphi}+\omega_i\sin\tilde{\varphi})^2+\omega_i^2\cos^2\tilde{\varphi}}{\left(\alpha_i^2+(\omega_i^2-\omega^2)\right)^2+4\alpha_i^2\omega^2}}$ , $\tilde{\varphi}=\varphi_{ki}+\psi_i$ , $\phi_{ki}$ is in the $k^{\text{th}}$ row and $i^{\text{th}}$ column of $\Phi$, $\psi_i$ is the $i^{\text{th}}$ column of $\Psi$, and $\lambda_i=\alpha_i+j\omega_i$ is the $i^{\text{th}}$ diagonal element of $\Lambda$.

Then, the expression of the standard deviation is derived. Remove the mean value from $I\%P_{a0}$ and define it as $\Delta P_{a0}=I\%(P_{a0}-\bar{P}_{a0})$. Let $\boldsymbol{u}(t)=\Delta P_{a0}\cos(\omega t+\varphi)$, the power spectrum of $\boldsymbol{u}(t)$ can be expressed as

$$S_{uu}(\Omega,t)=\sigma^2\cos^2(\omega t+\varphi)/W \quad (9)$$

where $\Omega$ represents the frequency variable, $\sigma$ is the standard deviation of $\Delta P_{a0}$ , and $W$ is the signal bandwidth.

Define a function $g(t)$ that satisfies $g(t)=|\cos(\omega t+\varphi)|$ for $t\geq 0$ and otherwise $g(t)=0$ . According to pseudo excitation method, by using $g(t)S_{uu}(\Omega)$ as incentives to (9), where $S_{uu}(\Omega)=\sigma^2/W$ , we can yield

$$\begin{aligned}\dot{\tilde{\boldsymbol{z}}}(\Omega,t)&=\boldsymbol{\Lambda}\tilde{\boldsymbol{z}}(\Omega,t)+\boldsymbol{\Psi}\tilde{\boldsymbol{u}}(\Omega,t)\\ \tilde{\boldsymbol{y}}(\Omega,t)&=\boldsymbol{\Phi}\tilde{\boldsymbol{z}}(\Omega,t)+\boldsymbol{D}\tilde{\boldsymbol{u}}(\Omega,t)\end{aligned} \quad (10)$$

where $\tilde{\boldsymbol{u}}(\omega,t)=g(t)\sqrt{S_{uu}(\Omega)}e^{j\Omega t}$ , $\tilde{\boldsymbol{z}}(\Omega,t)$ and $\tilde{\boldsymbol{y}}(\Omega,t)$ are pseudo variables, $\tilde{y}_k(\Omega,t)=\sum_{i=1}^{N}\phi_{ik}\psi_i\int_{-\infty}^{t}e^{\lambda_i(t-\tau)}g(\tau)\sqrt{S_{uu}(\Omega)}e^{j\Omega\tau}d\tau$ $=\sum_{i=1}^{N}\frac{\sigma}{\sqrt{W}}\phi_{ik}\psi_i H_i(\Omega,t)$, and $H_i(\Omega,t)=\int_0^t e^{\lambda_i(t-\tau)}|\cos(\omega t+\varphi)|e^{j\Omega\tau}d\tau$ .

Based on (10), the power spectrum of $\tilde{y}_k(\Omega,t)$ is

$$\begin{aligned}S_{y_k y_k}(\Omega,t)&=\tilde{y}_k^*(\Omega,t)\tilde{y}_k^T(\Omega,t)\\ &=\frac{\sigma^2}{W}\left(\sum_{i=1}^{N}\phi_{ik}^*\psi_i^*H_i^*(\Omega,t)\right)\left(\sum_{j=1}^{N}\phi_{kj}\psi_j H_j(\Omega,t)\right)^T\end{aligned}$$

Thus, the standard deviation of $y_k$ can be obtained by using

$$\sigma_{y_k}^2(t)=\int_{-\infty}^{\infty}S_{y_k y_k}(\Omega,t)d\Omega \quad (11)$$

Main components of $\sigma_{y_k}$ include the square root of the sum of the squares (SRS) term and the complex quadratic combination (CQC) term, which are given below.

1) The SRS term is

$$\begin{aligned}\sigma_{y_k\mathrm{SRS}}^2(t)&=\phi_{ik}^*\psi_i^*\mathrm{H}_i^*(\Omega,t)\mathrm{H}_j(\Omega,t)\psi_j^T\phi_{kj}^T\\ &=|\Phi_{ki}|^2|\Psi_i|^2\left|\frac{e^{j(\omega+\Omega)t+j\varphi}-e^{\lambda_i t+j\varphi}}{j(\omega+\Omega)-\lambda_i}-\frac{e^{-j(\omega-\Omega)t-j\varphi}-e^{\lambda_i t-j\varphi}}{j(\omega-\Omega)+\lambda_i}\right|^2\end{aligned} \quad (12)$$

When $\Omega\text{-}\omega_i+\omega=0$ or $\Omega\text{-}\omega_i\text{-}\omega=0$ , (12) can be given by

$$\sigma_{y_k\mathrm{SRS}}^2(t) \approx \frac{\sigma^2}{2W}\left|\Phi_{ki}\right|^2\left|\Psi_i\right|^2\sum_{i=1}^{n-1}\frac{1}{\alpha_i^2}(1-e^{\alpha_i t})^2 . \tag{13}$$

2) The CQC term is

$$\sigma_{y_k,CQC}^2(t) = \frac{\sigma^2}{W}\int_W\sum_{i=1}^{n}2\,\mathrm{Re}\left(\Phi_{ik}^*\Psi_i^*I_i^*(\Omega,t)I_j(\Omega,t)\Psi_j^T\Phi_{kj}^T\right)d\Omega \cdot \tag{14}$$

When $\Omega=0$, $\omega_i-\omega_j=2\omega$ and $i,j\in\Xi$, (14) can be given by

$$\sigma_{y_k,CQC}^2(t) \approx \frac{\sigma^2}{2W}\left|\xi_{ij}\right|\sum_{\Xi}\frac{1}{\alpha_i^2}(1-e^{\alpha_i t})^2\cos(2\omega t-\angle\xi_{ij}) \tag{15}$$

where $\xi_{ij}=\Phi_{ki}^*\Psi_i^*\Psi_j\Phi_{kj}$.

At this point, formulas for the mean and standard deviation of the simulated forced oscillation have been derived. Based on (8-15), the following five laws are concluded to describe the characteristics of the simulated forced oscillation.

***Law* 1**: The MMA response consists of the mean response (8) and the stochastic response (11). The mean response is very large under the influence of the frequency resonance. And the stochastic response is small under the excitation of wideband random MMA signals.

***Law* 2**: The mean response is composed of the free oscillation and the resonance oscillation whose frequency is the same as that of the disturbance. The weaker the inherent mode damping ratio, the greater the oscillation amplitude.

***Law* 3:** The closer the frequency of the MMA attack signal is to the inherent mode, the greater the forced oscillation amplitude, and the closer of the mode shape.

***Law* 4**: The beat frequency oscillation could be observed when the frequency of the MMA attack signal is different from the frequency of the free oscillation.

***Law* 5**: The variance of the stochastic response mainly includes the DC component, and the component whose frequency is twice the MMA attack signal.

It can be inferred from above laws that the simulated forced oscillation is different from free oscillation, which is easy to recognize [17]. For the suppression of the stimulated forced oscillation by MMA, measures can be taken from two aspects: the adjustment of the electromechanical mode and the suppression of disturbance source. On the one hand, we can increase the damping of the power system, or quickly control the natural frequency of the power system away from the MMA frequency, such as power system stabilizer. On the other hand, we can reduce the intensity of the disturbance source through isolating charging piles attacked by MMA or actively compensating the MMA disturbance.

## IV. The Defense Strategy against MMA

Designing a suitable controller to defend against MMA is necessary, however, it is a tricky problem. As an open ecosystem for the plug-in devices, CCLCS contains many system vulnerabilities. Moreover, different IoT devices have different standard protocols and system programs. And the same equipment produced by different manufacturers are also different, e.g., user App, communication facility, vehicle system, cloud system, etc. Thus, it is hard for the power system to design a dedicated defense grogram for each device. And it is also difficult for the cyber system to completely avoid all cyber-attacks, because a perfect cyber defense system for all attacks has not yet been proposed. In addition, isolating the charging piles attacked by MMA from the power system may be a solution. However, even if the attacked charging substation is detected by the forced oscillation analysis method, it is still hard to locate the exact charging pile quickly and accurately. Hence, the entire charging station has to be isolated, which will cause a large number of normal charging piles to be out of service. This is not an economical solution.

Based on the conclusion of part B in Section III, this section aims to propose a defense approach that passively strengthens the damping of the attacked electromechanical mode and actively compensates MMA disturbance to promote appropriate defensive reactions.

Usually, MMA is unexpected, setting compensation values for power disturbances injected by MMA is hard to accomplish in advance. Based on this consideration, active disturbance rejection control is very promising since it can provide compensation in real time for injected disturbance [18]. However, as the relative order of control is unable to determined, traditional active disturbance rejection control is difficult to directly apply to the power system. And the stability of ADRC cannot be guaranteed. Hence, there is a risk of instability when using this method.

To deal with the above problem, this section proposes a new multi-index information active disturbance rejection control (MIADRC) for the defense of MMA, and also gives the stability analysis of MIADRC.

### *A. Multi-index Information Active Disturbance Rejection Control (MIADRC)*

Assume that there are *m* machines in the power system. And each machine includes a third-order generator model and a first-order static exciter model [19]. Then, one can build the following state space expressions [19] for the *m* machines

$$\begin{bmatrix}\dot{\delta}_i\\ \dot{\omega}_i\\ \dot{E}_{qi}'\\ \dot{E}_{fi}\end{bmatrix}=\begin{bmatrix}\omega_i-\omega_0\\ \frac{\omega_0}{T_{Ji}}\left(P_{mi0}-P_{ei}\right)-\frac{D_i}{T_{Ji}}\left(\omega_i-\omega_0\right)\\ -\frac{1}{T_{d0i}'}\left(E_{qi}'+\left(x_{di}-x_{di}'\right)I_{di}\right)+\frac{1}{T_{d0i}'}E_{fi}\\ -\frac{1}{T_{Ai}}E_{fi}+\frac{k_{Ai}}{T_{Ai}}\left(u_{ref}-u_{ti}\right)\end{bmatrix}+\begin{bmatrix}0\\0\\0\\ \frac{k_{Ai}}{T_{Ai}}\end{bmatrix}u_{ei} \tag{16}$$

where $i=1,2,\cdots,m$.

Stimulated by MMA, forced oscillations of the attacked electromechanical mode will appear among *m* generators. The proposed MIADRC is a decentralized strategy, which provides additional control signals to *m* exciters and suppresses the MMA forced oscillation in every generator.

The key to construct the MIADRC defense strategy is to design a suitable input signal. We propose a multi-index information form for the input *y*(*t*) of MIADRC based on the idea of multi-index nonlinear coordinated control [20], which can be expressed as

$$y_i=h_i(X)=c_{1i}\Delta u_{ti}+c_{2i}\Delta\omega_i+c_{3i}\Delta P_{ei} \tag{17}$$

where $c_{1i}=1$, $c_{2i}=-0.1$, $c_{3i}=0.5$.

To calculate the relative order of (17), we employ the differential geometric linearization method. By combining (16) and (17), we construct an $m$-input $m$-output nonlinear $m$-machine power system expressed as

$$\dot{X}(t)=f(X(t))+g(X(t))u(t), \quad y(t)=h(X(t)), \tag{18}$$

where $u(t)\in R^m$ is the input of $m$ exciters, and $y(t)\in R^m$ represents $m$ output signals of the $m$-input $m$-output system. Let $G=\{g_1,g_2,\cdots,g_i,\cdots,g_n\}$ denote the distribution generated by $g_i\in g(X(t))$, where Lie bracket for any $g_i$ and $g_j$ is zero $[g_i,g_j]=0$ and G obeys the involutive distribution. Based on Frobenius theory, (18) can be transformed to the following normal model by selecting $z_{i1}=y_{i1}$, $z_{i2}=\dot{y}_{i2}$, $z_{i3}=\delta_i$, and $z_{i4}=\omega_i$

$$\begin{aligned} z_{i1}&=z_{i2}\\ z_{i2}&=L_f^2 z_{i1}+\textstyle\sum_{j=1}^{m} L_{gj}L_f z_{i1}u_e\\ z_{i3}&=z_{i4}\\ z_{i4}&=q_i(z_{i1},z_{i2},z_{i3}) \end{aligned} \tag{19}$$

where, $L_{gj}L_f^0 z_{i1}=0$ , $L_{gj}L_f^1 z_{i1}=\dfrac{c_{1i}K_{Aj}}{T'_{d0j}T_{Aj}}\dfrac{\partial u_{ti}}{\partial E'_{qj}}+\dfrac{c_{3i}K_{Aj}}{T'_{d0j}T_{Aj}}\dfrac{\partial P_{ei}}{\partial E'_{qj}}$ ,.

Expressions of $L_f^2 z_{i1}$ and $q_i(z_{11},z_{12},z_{13},z_{14},z_{21},\cdots,z_{m4})$ are neglected to save space since they do not affect the design process of MIADRC.

From (19), it can be concluded that the relative order from input to output is two. Then, the MIADRC defense strategy is constructed based on (19) and second-order active disturbance rejection control. It consists of multi-index information input nonlinear tracking-differentiator (TD), extend state observer (ESO) and nonlinear state feedback (NSF) output. The resulted algorithm of the MIADRC controller are as follows

$$TD:\begin{cases} fh=fhan(v_{i1}(k)-r_i,v_{i2}(k),r_0,h)\\ v_{i1}(k+1)=v_{i1}(k)+hv_{i2}(k)\\ v_{i2}(k+1)=v_{i2}(k)+hfh \end{cases} \tag{20-a}$$

$$ESO:\begin{cases} e(k)=\chi_{i1}(k)-y_i(k)\\ \chi_{i1}(k+1)=\chi_{i1}(k)+h(\chi_{i2}(k)-\beta_{i1}e(k))\\ \chi_{i2}(k+1)=\chi_{i2}(k)+h(\chi_{i3}(k)-\beta_{i2}e(k)+u_i(k))\\ \chi_{i3}(k+1)=\chi_{i3}(k)+h(-\beta_{i3}e(k)) \end{cases} \tag{20-b}$$

$$NSF:\begin{cases} e_{i1}=v_{i1}(k+1)-\chi_{i1}(k+1)\\ e_{i2}=v_{i2}(k+1)-\chi_{i2}(k+1)\\ u_i(k+1)=-fhan(e_{i1},ce_{i2},r_0,h)-\chi_{i3}(k+1)\\ u_{ei}(k+1)=b_i^{-1}u_i(k+1) \end{cases} \tag{20-c}$$

where *fhan* adopts the function in [18], $h$=0.01, $c$=0.5, $r_i$=0, $r_0$=0.01, $\beta_{i1}=3w_c$ , $\beta_{i2}=3w_c^2$ , $\beta_{i3}=w_c^3$ , $w_c$=100$\pi$ .

During the design process of MIADRC, one critical part is to design the value of the parameter $b_i$. It can be derived from (19) that $b_i$ is $b_i=L_{gi}L_f^1 z_{i1}=c_{1i}\dfrac{K_{Ai}}{T'_{d0i}T_{Ai}}\dfrac{\partial u_{ti}}{\partial E'_{qi}}+c_{3i}\dfrac{K_{Ai}}{T'_{d0i}T_{Ai}}\dfrac{\partial P_{ei}}{\partial E'_{qi}}$ . Let $b_i$ take the initial value $b_i=b_i(0)$ , and consider the error introduced as disturbance which can be compensated by ESO. Then, ESO observes the model error and the MMA disturbance (19), which are dynamically compensated by NSF. And only local variables need to be measured. The control logic of MIADRC on the $i^{th}$ generator is shown in Fig. 4 where (a) provides the structure of MIADRC and (b) corresponds to the control flow chart. An online identification program is run to detect MMA. Once MMA is detected, the switch S1 in Fig. 4 (a) is closed immediately, which directly put MIADRC into operation. Then, the output of MIADRC compensates the disturbance and suppresses the forced oscillation brought by MMA. Finally, when MMA stops, S1 is open to isolate MIADRC from the power system.

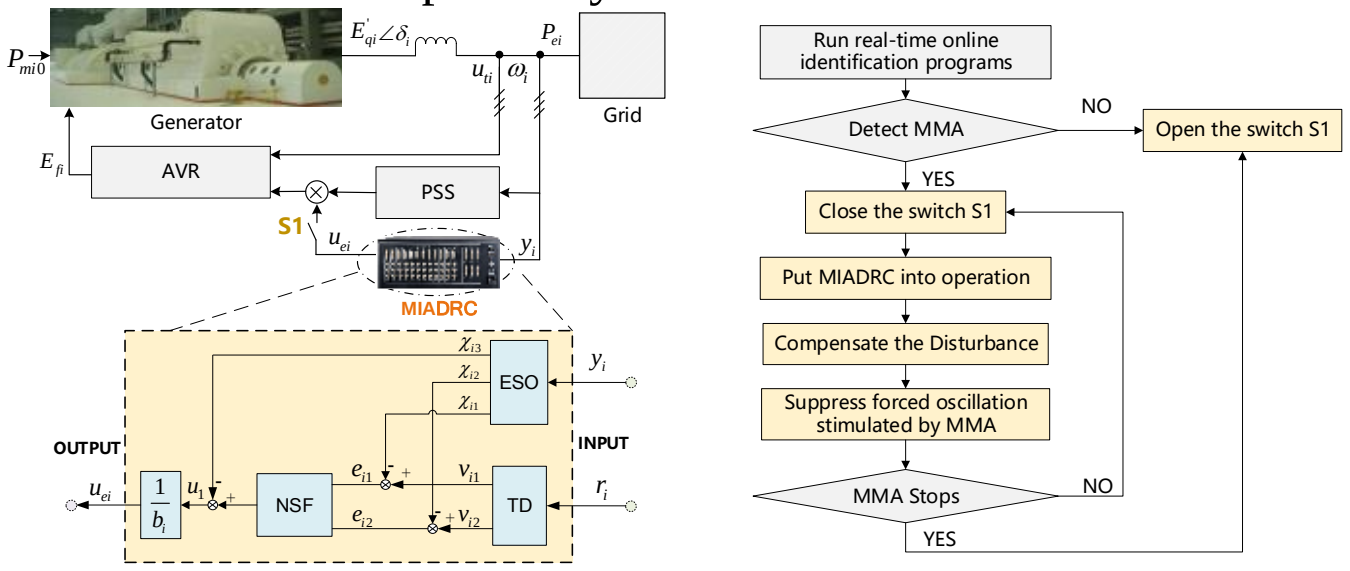

(a) MIADRC (b) The control flow chart of MIADRC

Fig. 4. The control logic of MIADRC.

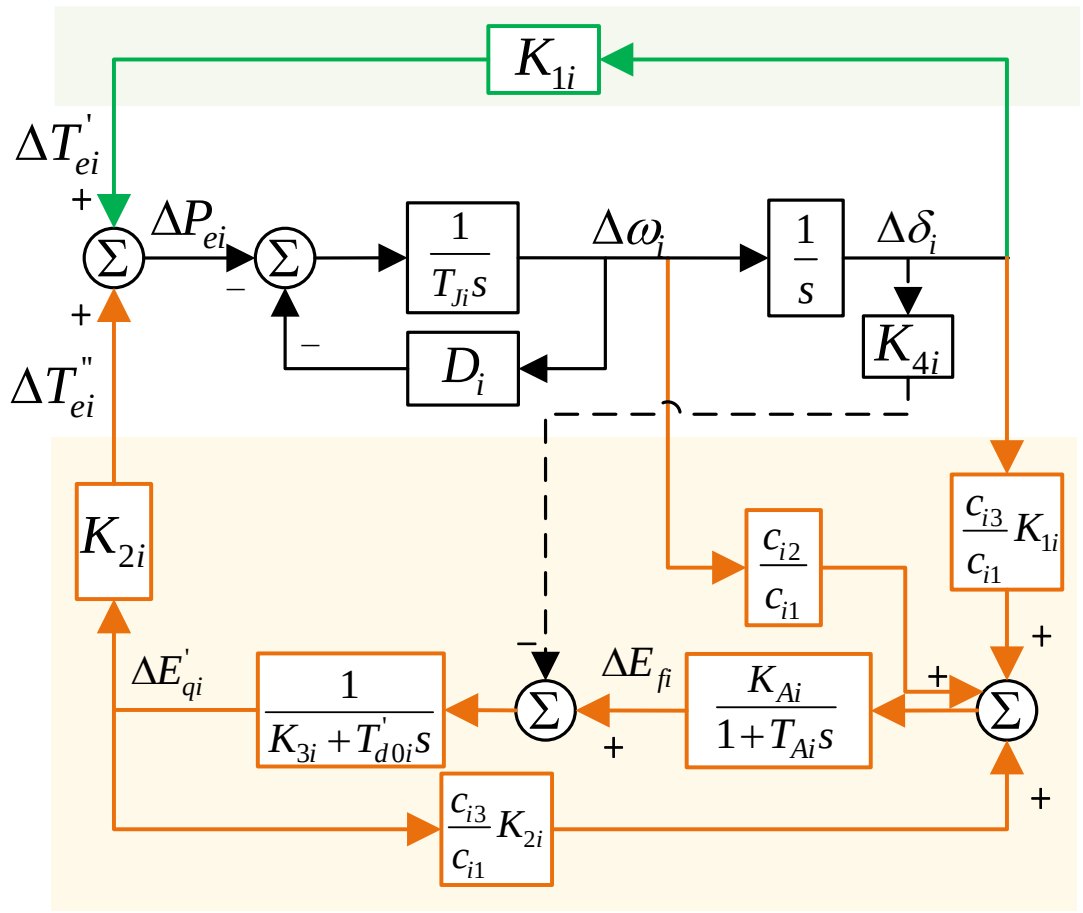

Fig. 5. Transfer function block diagram under zero dynamic.

## *B. Stability Analysis on the Proposed MIADRC Strategy*

This part mainly analyzes the stability of the proposed MIADRC strategy. Considering that MIADRC can effectively compensate the model and parameter error by ESO and NSF, the zero-dynamic equations of (19) for each generator are

$$\dot{\delta}_i=\omega_i-\omega_0 \tag{21-a}$$

$$T_{Ji}\dot{\omega}_i=P_{mi0}-P_{ei}-D_i(\omega_i-\omega_0) \tag{21-b}$$

$$T'_{d0i}\dot{E}'_{qi}=-(E'_{qi}+(x_{di}-x'_{di})I_{di})+E_{fi} \tag{21-c}$$

$$T_{Ai}\dot{E}_{fi}=-E_{fi}-K_{Ai}\left(-\frac{c_{2i}}{c_{1i}}\Delta\omega_i-\frac{c_{3i}}{c_{1i}}\Delta P_{ei}\right) \tag{21-d}$$

By linearizing (21), we obtain the linearized system model (A1) in Appendix A, and its transfer function block diagram is shown in Fig. 5. Based on Fig. 5, the damping torque except $D_i$ equals

$$T_{ei}=\Delta T'_{ei}+\Delta T''_{ei}=K_{1i}\Delta\delta_i+\frac{K_{2i}K_{Ai}\left(\frac{c_{i3}}{c_{i1}}K_{1i}\Delta\delta_i+\frac{c_{i2}}{c_{i1}}s\Delta\delta_i\right)}{(1+T_{Ai}s)(K_{3i}+T'_{d0i}s)-\frac{c_{i3}}{c_{i1}}K_{2i}K_{Ai}} \tag{22}$$

where $K_{1i}=\frac{\partial P_{ei}}{\partial \delta_i}$, $K_{2i}=\frac{\partial P_{ei}}{\partial E_{qi}^{'}}$, $K_{3i}=\frac{\partial E_{qi}}{\partial E_{qi}^{'}}$, and $K_{4i}=\frac{\partial E_{qi}}{\partial \delta_i}$.

Let $s=jw$ , (22) can be rewritten as

$$T_{ei}=K_{ei}\Delta\delta_i+D_{ei}\Delta\omega_i \tag{23}$$

with $D_{ei}=\dfrac{K_{2i}K_{Ai}\left[\frac{c_{i2}}{c_{i1}}(-K_{Ai}\frac{c_{i3}}{c_{i1}}K_{2i}+K_{3i}-w^2T_{Ai}T_{d0i}^{'})-\frac{c_{i3}}{c_{i1}}K_{1i}(K_{3i}T_{Ai}+T_{d0i}^{'})\right]}{(-K_{Ai}\frac{c_{i3}}{c_{i1}}K_{2i}+K_{3i}-w^2T_{Ai}T_{d0i}^{'})^2+w^2(K_{3i}T_{Ai}+T_{d0i}^{'})^2}$,

$$K_{ei}=K_{1i}+K_{2i}K_{Ai}\frac{\frac{c_{i3}}{c_{i1}}K_{1i}(-K_{Ai}\frac{c_{i3}}{c_{i1}}K_{2i}+K_{3i}-w^2T_{Ai}T_{d0i}^{'})+w^2\frac{c_{i2}}{c_{i1}}(K_{3i}T_{Ai}+T_{d0i}^{'})}{(-K_{Ai}\frac{c_{i3}}{c_{i1}}K_{2i}+K_{3i}-w^2T_{Ai}T_{d0i}^{'})^2+w^2(K_{3i}T_{Ai}+T_{d0i}^{'})^2}.$$

Usually, the excitation regulator parameters satisfy $T_{Ai}\approx 0$, and $K_{Ai}\geq 1$ . Therefore, it can be concluded that the electromagnetic torque will provide the positive damping $D_{ei}>0$ if $c_{i3}c_{i2}<0$ and $c_{i1}>0$.

Based on the above analysis, we set $c_{i2}c_{i3}=-0.05$, and $c_{i1}=1$ to ensure the stability of the designed MIADRC when defending against MMA.

## V. Case Simulation and Study

In this section, a test platform was built to verify the MMA threat and verify the mathematical model for MMA analysis. And characteristics of MMA are also demonstrated, followed by the validation of the effectiveness of the proposed MIADRC defense strategy.

### *A. Verify the MMA Threat and the Model for MMA Analysis*

A hardware test platform, as shown in Fig.6, is established to verify the MMA threat and the mathematical model for MMA analysis. To build this platform, we connected a coordinated charging pile to a power source that simulates the infinite bus, used Raspberry Pi with FreeRTOS system installed to imitate an energy router, and adopted AliBaba Cloud server to imitate the charging cloud which is connected to the Raspberry Pi (energy router) using the MQTT protocol. We assumed that the charging control cloud can be infected by the MMA Trojans, which was simulated by presetting the MMA malicious program. This program was designed to downloaded into the Raspberry Pi via the cloud service, and then manipulate the output power of the charging pile. We are also equipped with a DL850EV oscillography, which can measure AC voltage and output power of the charging pile during the period of MMA.

Based on this test platform, we first verified the correctness of the proposed charging pile model for MMA analysis. As MMA attacks the active power command of the charging pile, the dynamic response to the change of $P_{\mathrm{ref}}$ is vital in constructing the model for MMA analysis. Thus, we perturbed the power command value in this test platform, and compared the measured dynamic response with the numerical simulation results of the proposed mathematical model.

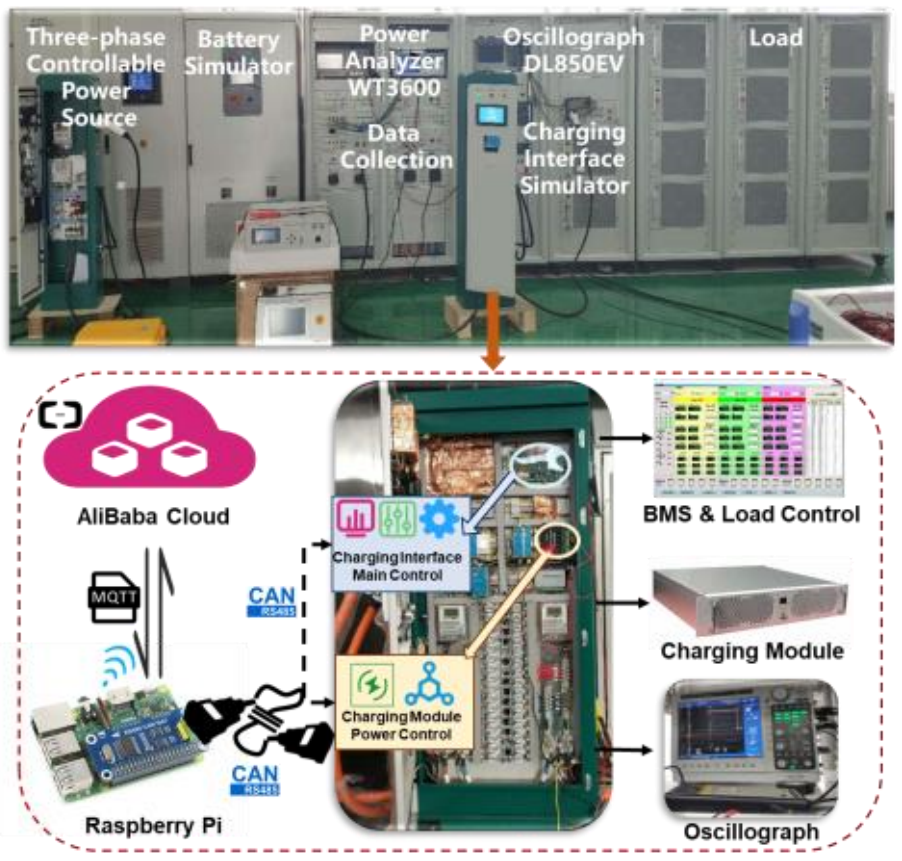

Fig. 6. The test platform for the MMA threat.

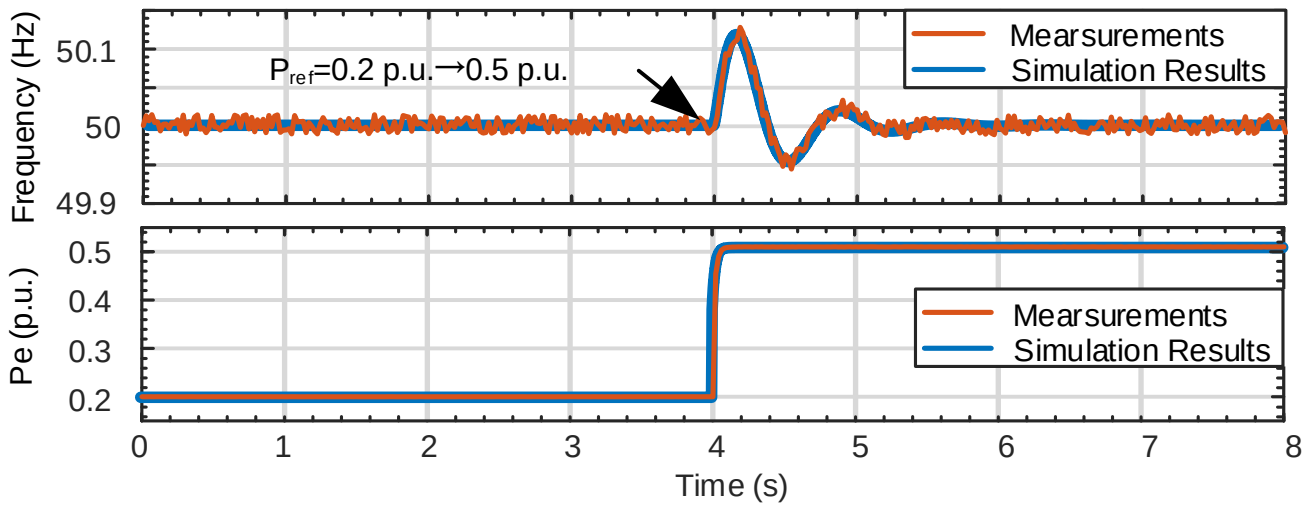

Fig. 7. Comparison of dynamic responses. The measurements by oscillography vs. simulated results in orange and blue line respectively.

The active power command of the charging pile in Fig. 6 was increased from 0.2 to 0.5 p.u, and the corresponding dynamic responses measured by the oscillography is shown as the orange line in Fig. 7. Meanwhile, by using the proposed model in Fig. 3 to model the charging pile, the time domain numerical simulation program also calculated the dynamic response of this disturbance, which is represented by the blue line in Fig. 7. It shows that the blue line completely reflects the electromechanical dynamics of the red line. Hence, the proposed charging pile model is applicable to the analysis of the forced oscillation caused by MMA which is an electromechanical-scale issue.

Moreover, the modal information in the measured and simulated dynamic curves of Fig. 7 is compared. The modal information of measured orange curve is identified by using Prony analysis. And the modal information of the simulated blue line is computed by the modal analysis program which adopts the linearized form of the proposed charging pile model (2). All obtained results are given in Table II. It shows that the error is lower than 1%. Therefore, the proposed charging pile model can reflect electromechanical-scale modal information with high accuracy.

TABLE II

Comparison of Model Information

| | Identified from Measurements | Computed Using MMA Model | Error |
|---|---|---|---|
| **Damp Ratio** | 0.2820 | 0.2797 | -0.82% |
| **Frequency (Hz)** | 1.363 | 1.370 | 0.51% |

Key parameters of the proposed charging pile model were also discussed. The participation factors of the charging pile state variables to the mode shown in Table II were calculated and shown in Fig. 8. It reveals that state variables $X_{pll}$ and $\theta_{pll}$ of PLL have greater influence. Then, we further computed the mode trajectory related to the parameters $k_{i3}$ and $k_{p3}$ adopted in PLL in Fig. 3. The obtained mode trajectories are shown in Fig. 9. It can be seen from Fig. 10 that $k_{p3}$ mainly affects the real part of this mode, and the mode decay rate increases as $k_{p3}$ increases. As for $k_{i3}$, it mainly affects the imaginary part of this mode, and, the mode frequency increases as $k_{i3}$ increases. Hence, for a large number of charging piles, the equivalent electromechanical model can be established by selecting appropriate $k_{p3}$ and $k_{i3}$. In the analyses on MMA, certain fixed parameters $k_{p3}$ and $k_{i3}$ can be assumed. And when the charging pile model is applied to a specific MMA instance, the corresponding parameter values can be determined based on the actual measurement data.

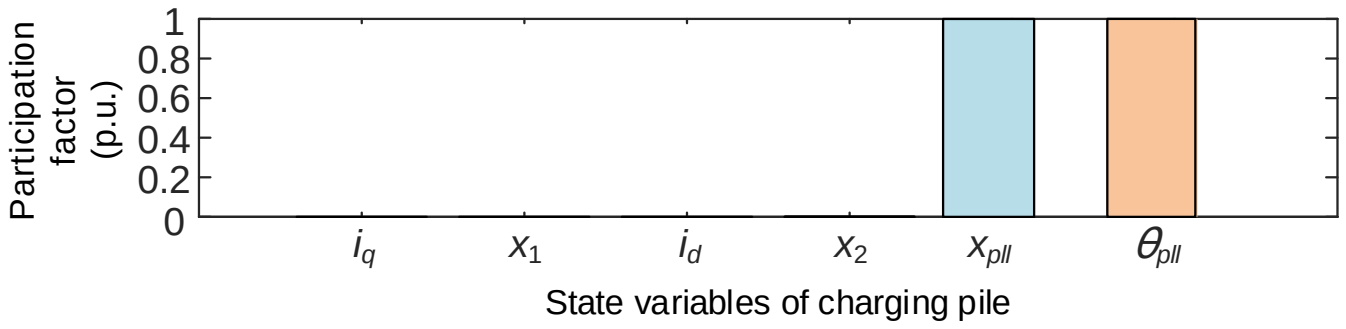

Fig. 8. The participation factor of charging pile for electromechanical mode.

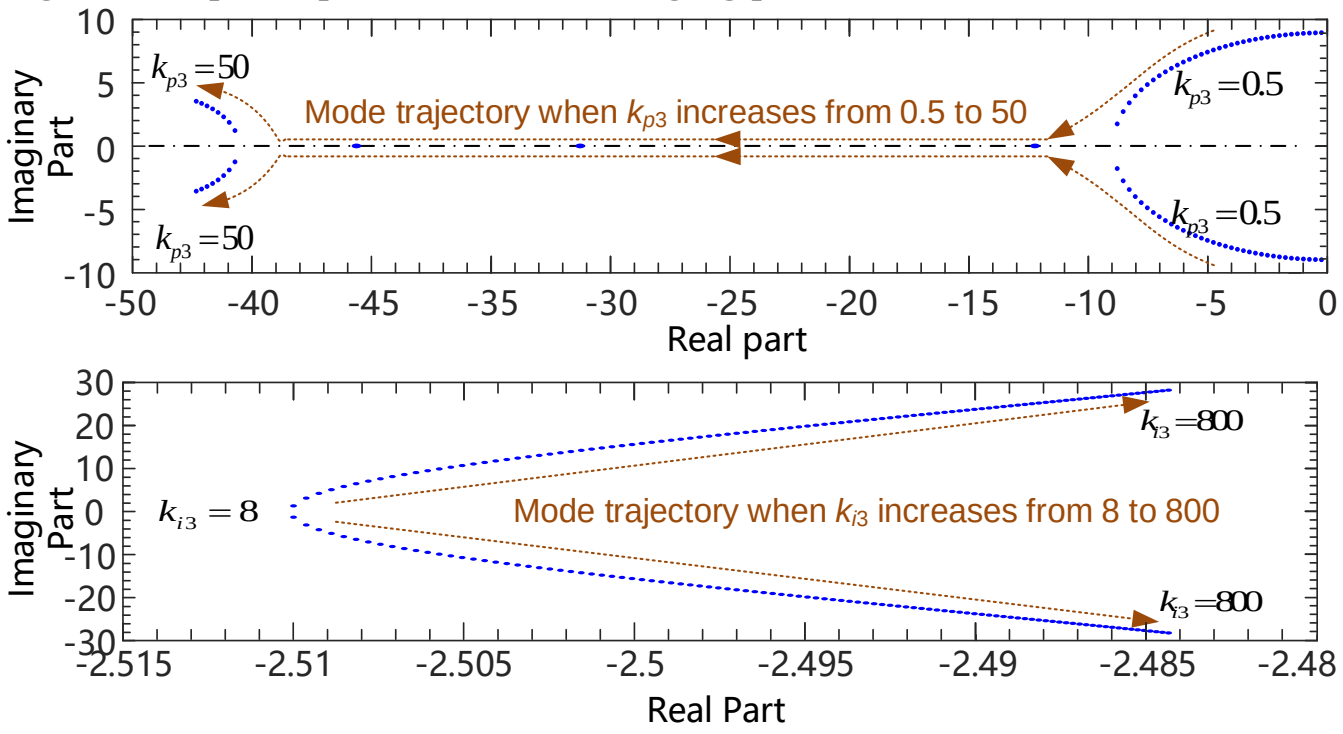

Fig. 9. Mode trajectories when $k_{p3}$ and $k_{i3}$ increase.

Next, we validated the MMA threat. As is shown in Table II, there is a 1.37Hz mode in the electrochemical time scale. Hence, we set the frequency $\omega$ of the MMA attack command (1) to 1.37Hz. Attacked by (1), the corresponding dynamic responses of active power of the charging pile were both measured in the test platform (Fig. 6) and simulated by the numerical simulation program, which generated the results in Fig. 10. It can be seen that the simulated dynamic response indicated by the blue line is quite close to the measured dynamic response indicated by the orange line. Moreover, both the measured and simulated results of the active power of the charging pile quickly track the MMA reference signal and oscillate. Hence, it is demonstrated that MMA can quickly cause obvious forced oscillation. Considering that in a bulk power system the forced oscillation usually affects two interconnected regional power grids, therefore, MMA is a great threat for the stability of the bulk power system.

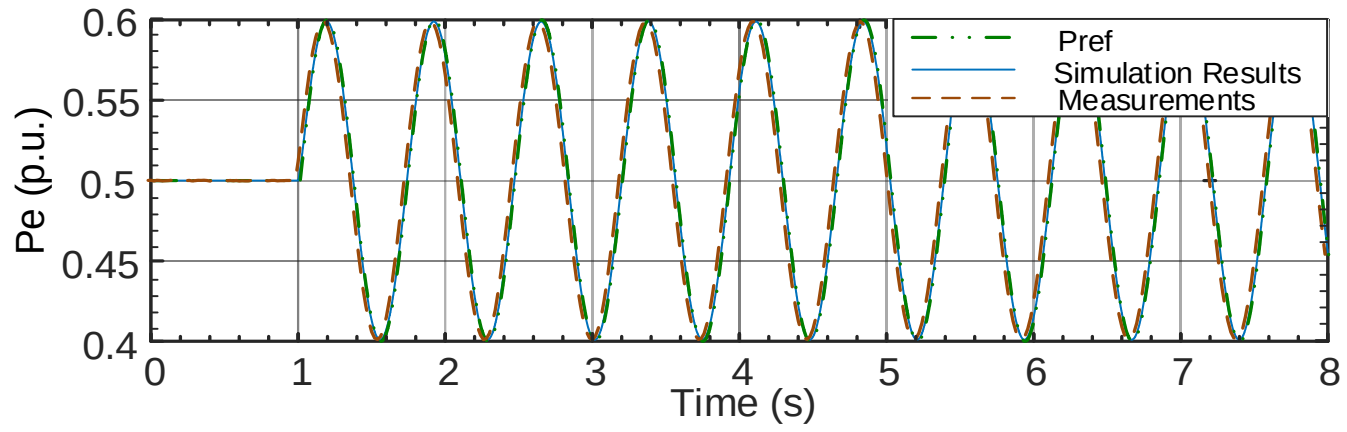

Fig. 10. The active power under MMA.

## B. Characteristics Test of Forced Oscillation Caused by MMA

In this section, characteristics of the stimulated forced oscillation dynamics as affected by MMA is tested and discussed by using IEEE 4-machine 2-area system [21] shown in Fig. 11. A large EV charging load is added at Bus 7 and adopts the charging pile model described by (2).

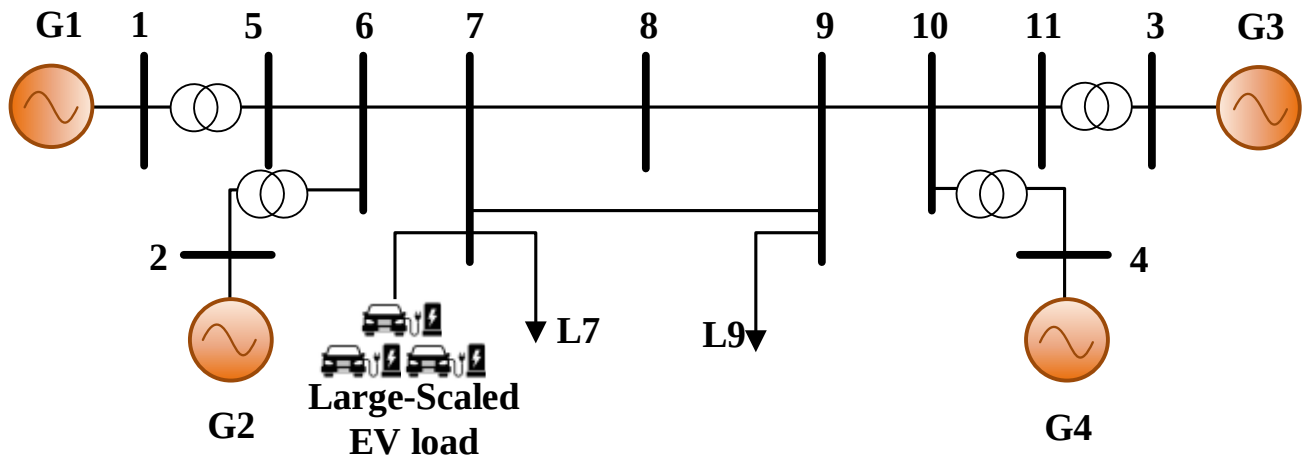

Fig. 11. The IEEE 4-machine 2-area test system with large-scaled EV load.

The impact of the power value of the EV charging load is analyzed. Increasing the charging load from 0.5 to 8 p.u., the mode trajectory of the wide-area oscillation mode related to (G1, G2) vs. (G3, G4) is computed and given in Fig. 12, where the corresponding damping and frequency are also calculated. It shows that the damp and the frequency first rise then fall. The varying range of damping ratio is from 0.074 to 0.022, and the varying range of frequency is from 0.46 to 0.62 Hz. Thus, the charging load has a great impact on the damping ratio of the wide-area oscillation mode.

In addition, the trajectory of the wide-area oscillation mode is also studied under different values of EV load within a day. A typical 24-hour coordinated charging load data was applied to the EV load in this test system (Fig. 11). The adopted charging load data and the corresponding damping and frequency of the wide-area oscillation mode are illustrated in Fig. 13. It shows that damping ratio is very low during 21:00~3:00 at night, which corresponds to the period near the peak value of EV charging load. This phenomenon is consistent with the varying trend of damping in the region of heavy EV load in Fig. 12. Therefore, the damping of the wide-area oscillation mode in this test system is weak during the period of heavy EV load.

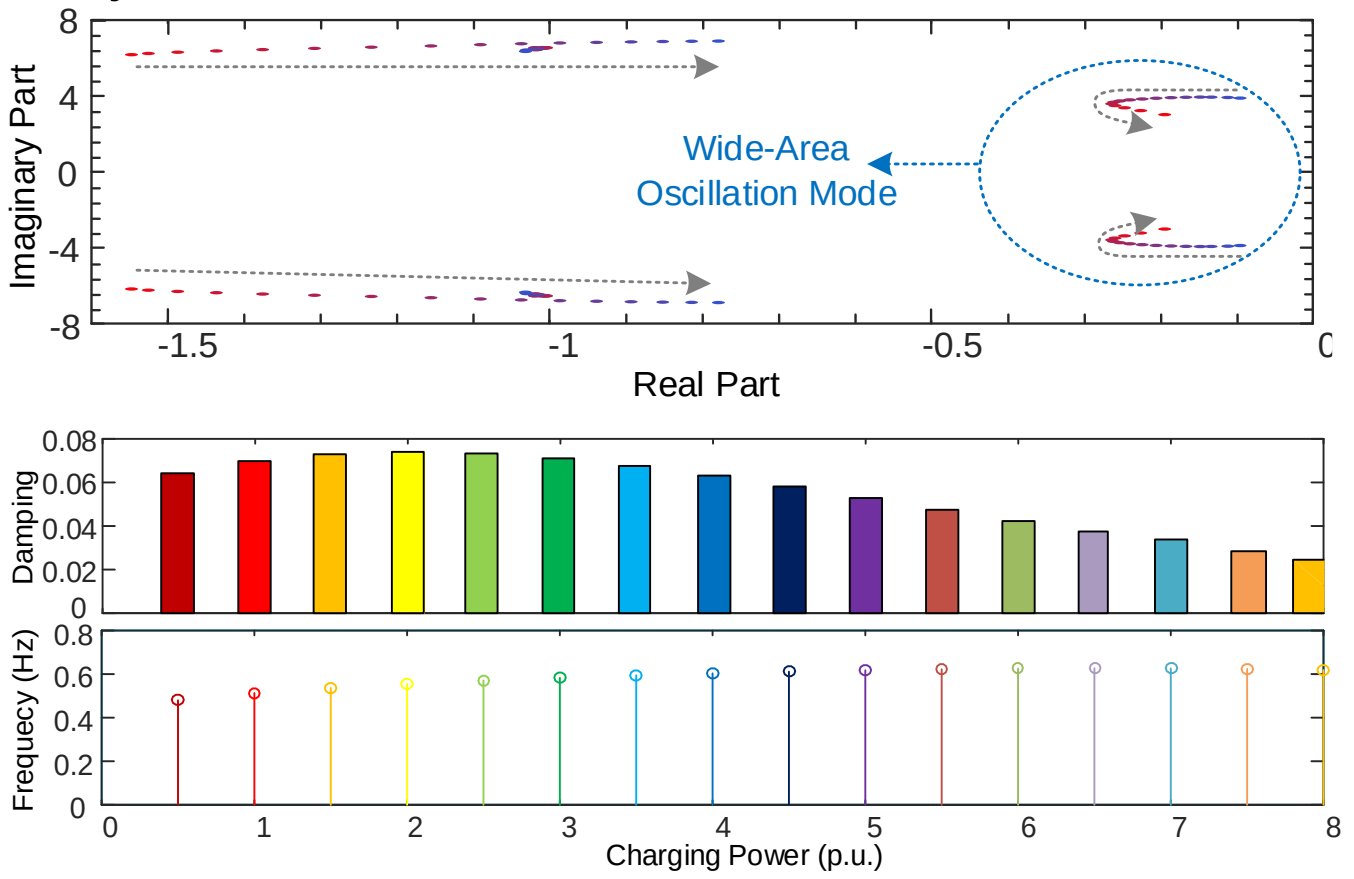

Fig. 12. The mode trajectory of the wide-area oscillation mode with increasing charging load, and the corresponding damping and frequency.

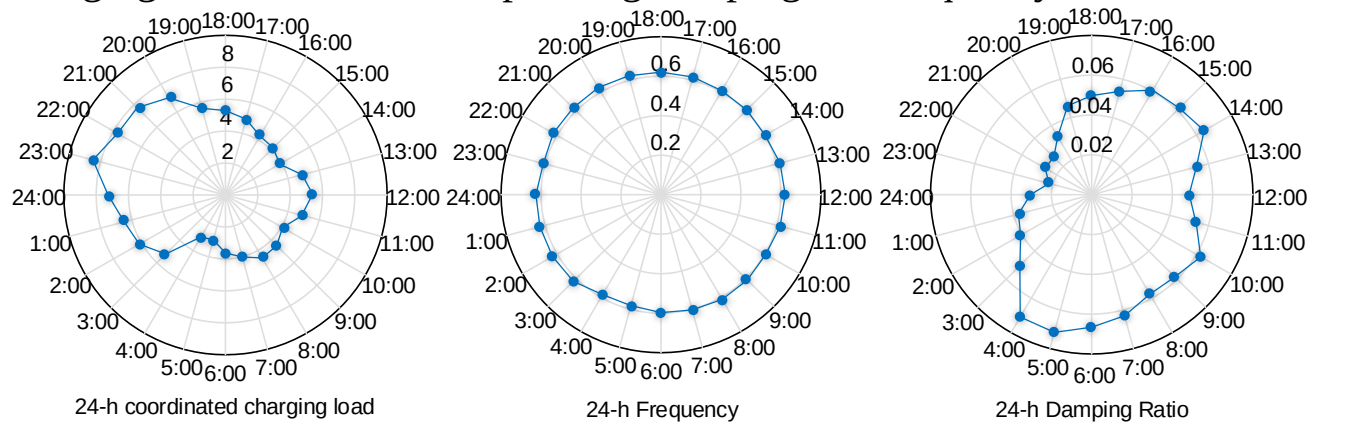

Fig. 13. Typical example of coordinated charging load with relevant frequency and damping.

Based on the tests above, heavy EV load greatly reduces the damping of the wide-area oscillation mode in this test system. As one can always detect a certain period near the charging peak of large-scaled EV load, setting the launch time for MMA is not a difficult task. Hence, this test system is greatly threatened by MMA. And in the following contents, this system is adopted to test the characteristics of the stimulated forced oscillation by MMA.

Summarizing Section III.B, we can draw the conclusion that the magnitude of the forced oscillation under MMA is affected by three factors: disturbance intensity, system damping and frequency deviation. These factors will be analyzed below.

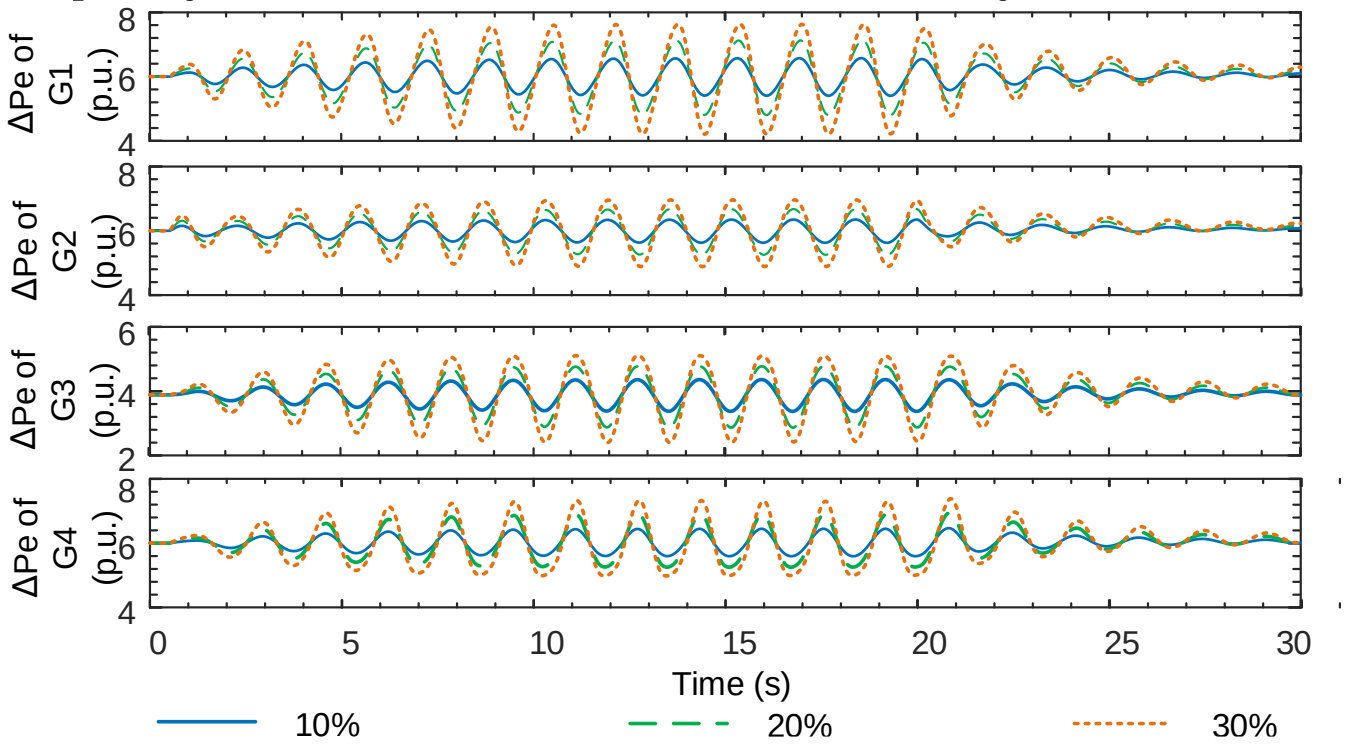

Fig. 14. Forced oscillations under different ratio I%.

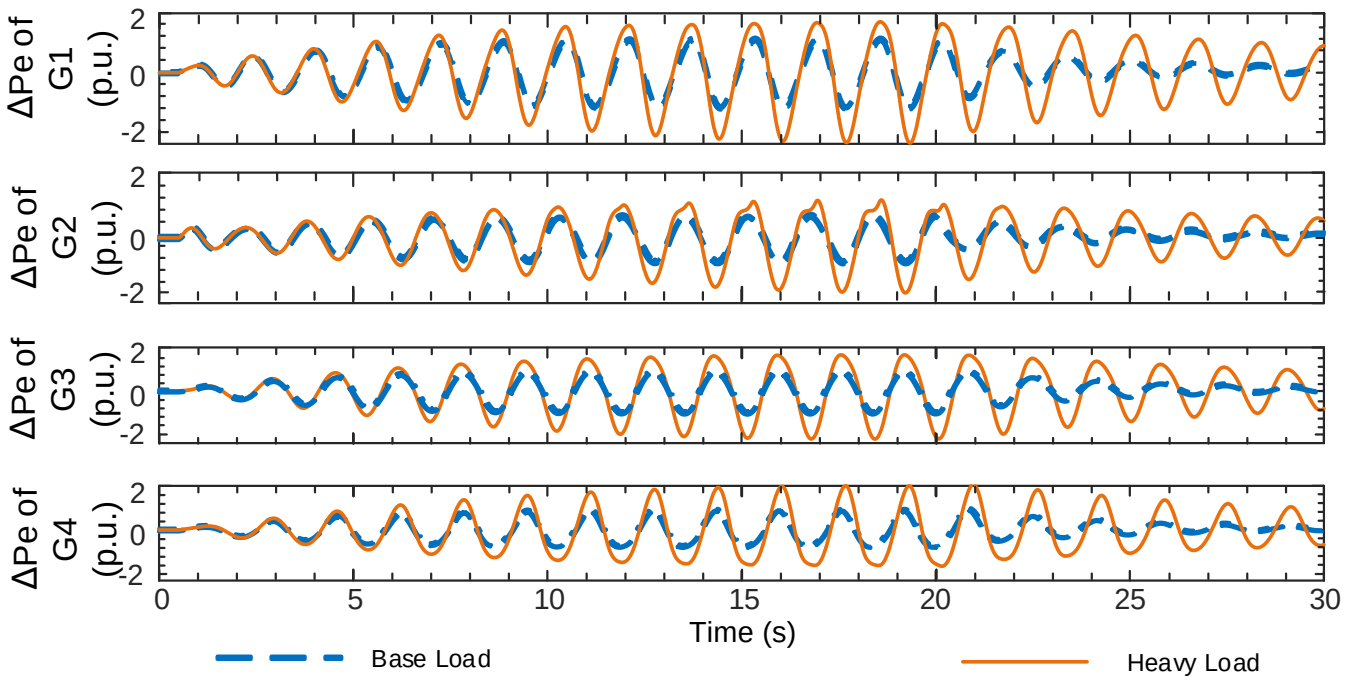

Fig. 15. Forced oscillations under different system damping conditions.

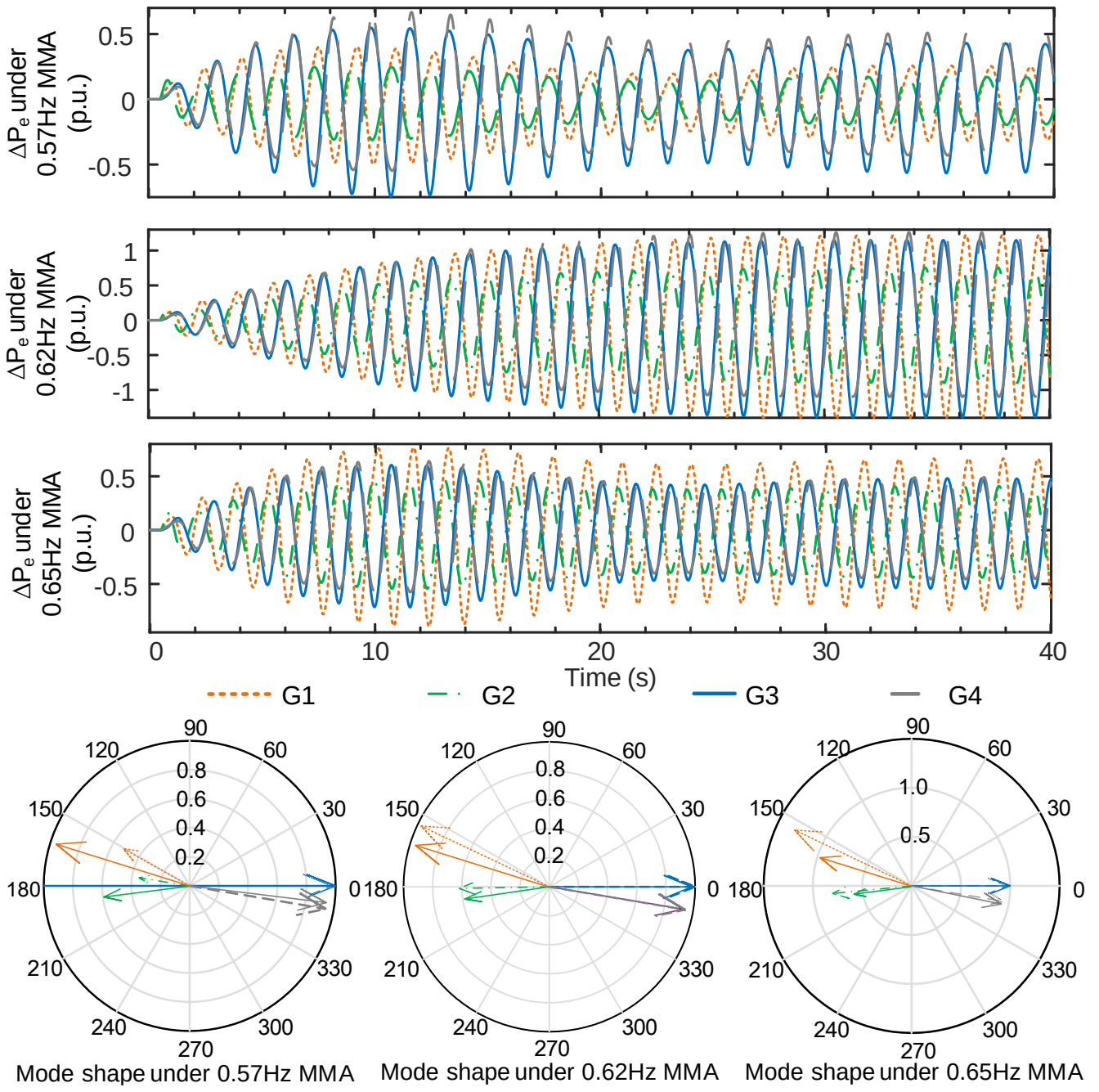

Fig. 16. Forced oscillations under different MMA frequency.

Firstly, the factor of disturbance intensity is adopted in the characteristic test. Since disturbance intensity is mainly affected by the amplitude of the MMA disturbance determined by disturbance ratio *I*% as defined in (1), simulations under different values of *I*% are conducted and shown in Fig. 14. It reveals that the oscillation amplitude raises as *I*% increases. As the mean response dominates the amplitude of the forced oscillation. And according to (7), the mean response is positively correlated with *I*%. Thus, this phenomenon that oscillation amplitude increases with *I*% is reasonable and in accordance with ***Law* 1**.

Next, the characteristic test mainly focuses on the factor of system damping. As demonstrated in Fig. 13, the damping ratio is greatly influenced by the EV load. Two scenarios of base EV load and heavy EV load are considered. The corresponding modal information is computed and listed in TABLE III, where the damping is sufficient under the base-load case and very weak under the heavy-load case. Simulations under both scenarios are conducted and presented in Fig. 15. It shows that the forced oscillation power amplitude in the heavy-load operation is significantly higher than that in the base-load operation, which infers that under MMA less damped system will lead to larger oscillation amplitude. Hence, this phenomenon demonstrates the correctness of ***Law* 2**.

TABLE III

DIFFERENT SYSTEM OPERATION CONDITIONS AND DAMPING RATIO

| Operation | L2 Load | Damp Ratio | Frequency(Hz) |
|---|---|---|---|
| **Base-Load** | 13.67 | 0.06 | 0.60 |
| **Heavy-Load** | 18.67 | 0.02 | 0.62 |

Further, the influence of the frequency is tested. The heavy-load case in TABLE III is adopted and different signal frequencies 0.57, 0.62 and 0.65 Hz are set for MMA. The corresponding responses of output power of four generators are given in Fig. 16, where the relevant mode shapes are also given. The 0.62 Hz MMA causes a forced oscillation with larger amplitude than that caused by 0.65 and 0.57 Hz. And the modal shape related to 0.62 Hz is closer to the inherent mode (dotted line), which is consistent to ***Law* 3**. Moreover, the beat frequency oscillations only appear in 0.65 and 0.57 Hz cases, which is in accord with ***Law* 4**.

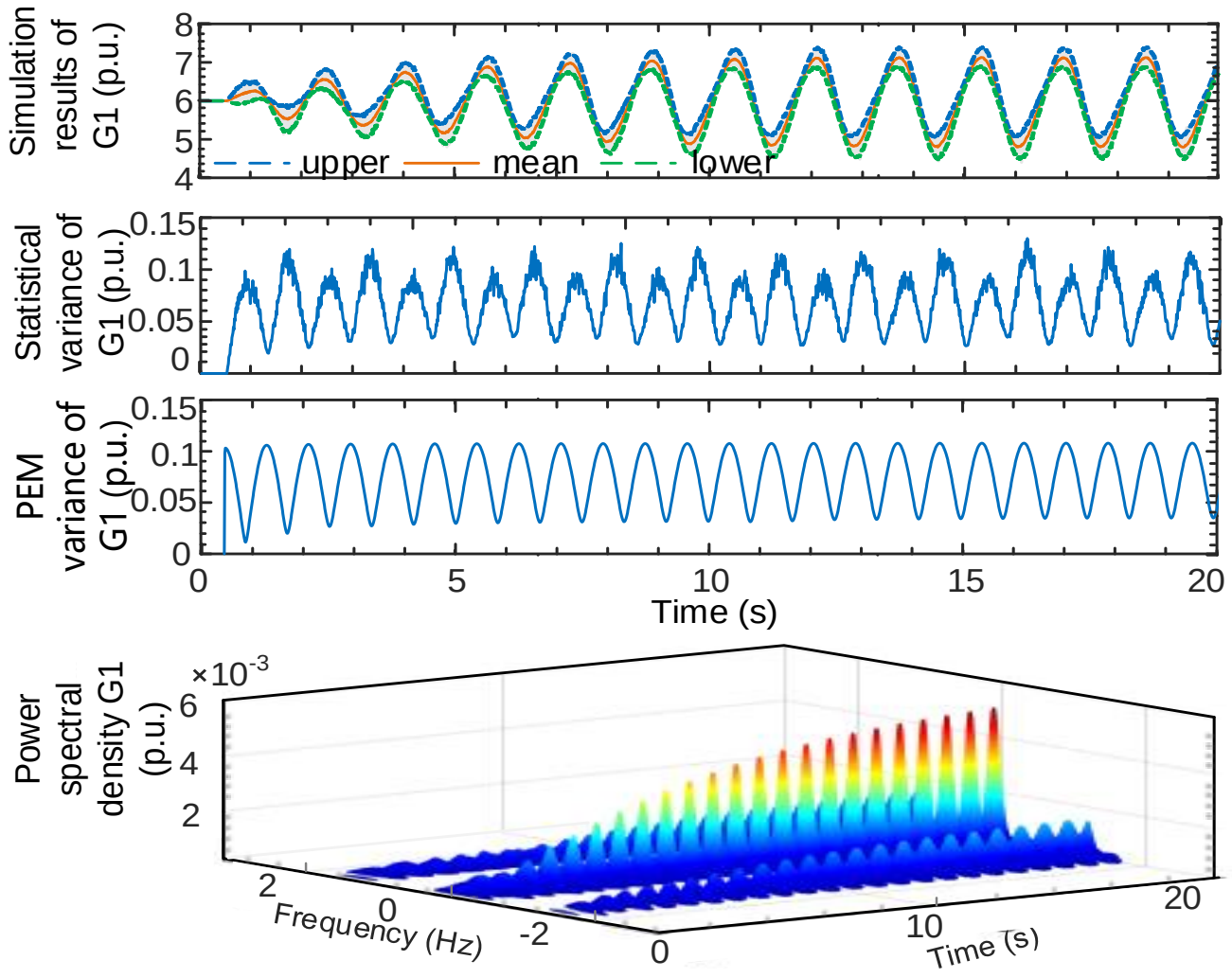

Fig. 17. Stochastic responses to $\Delta P_{a0}$ =2.4p.u., with the Monte Carlo simulation and its statistical variance, statistical variance given by PEM method, and power spectral density.

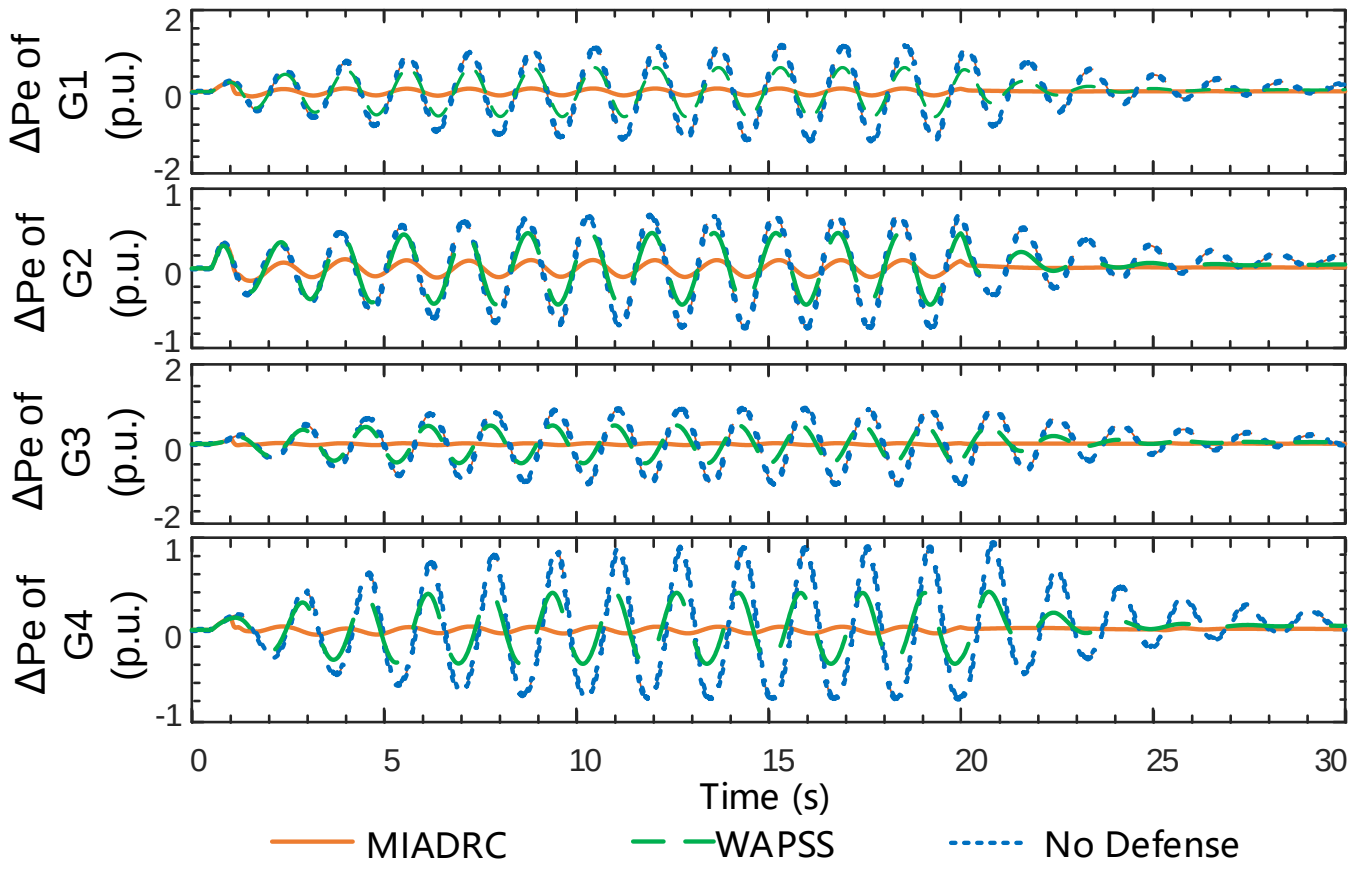

Fig. 18. The MIADRC performance under the basic condition.

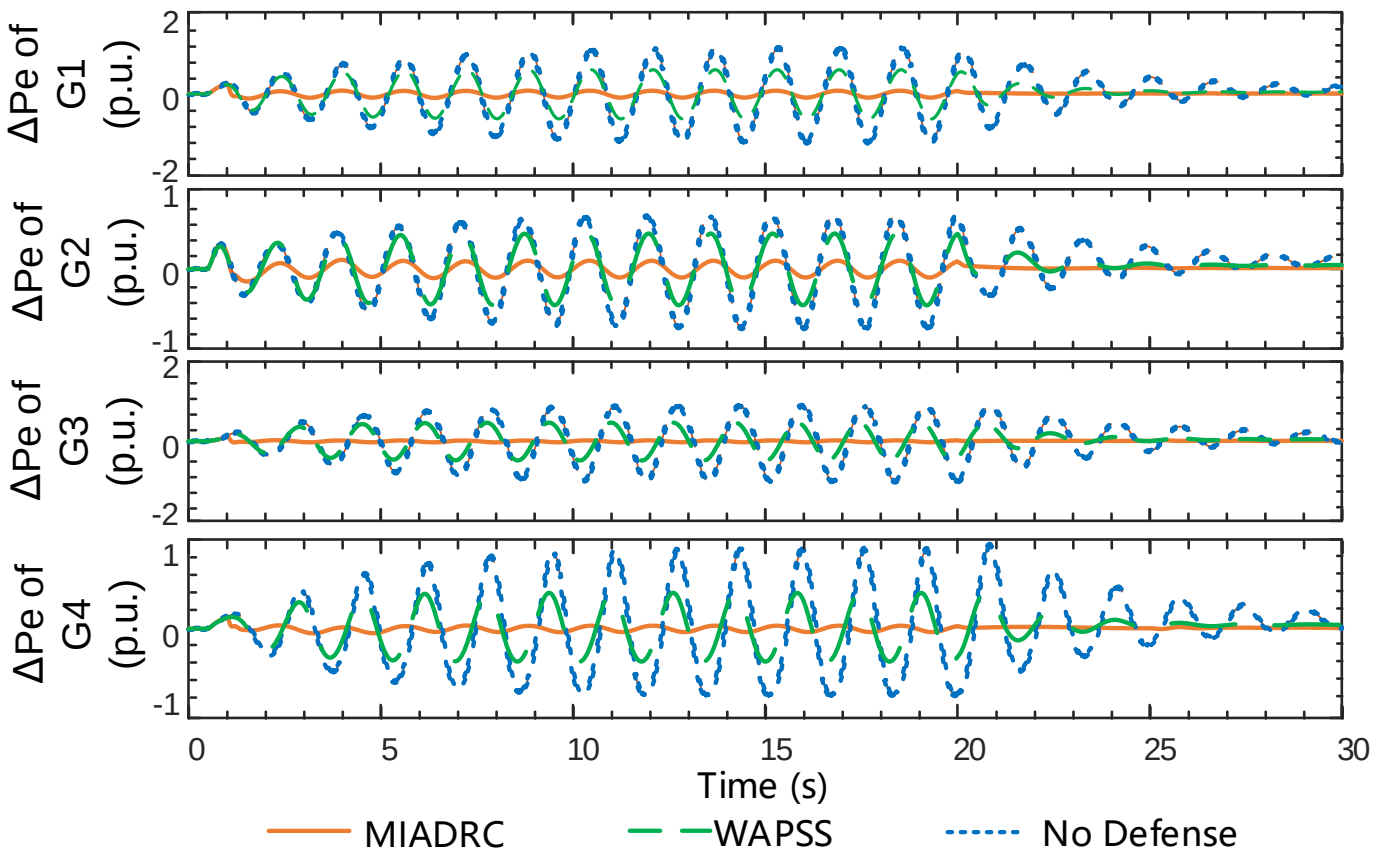

Fig. 19. The MIADRC performance under the tense condition.

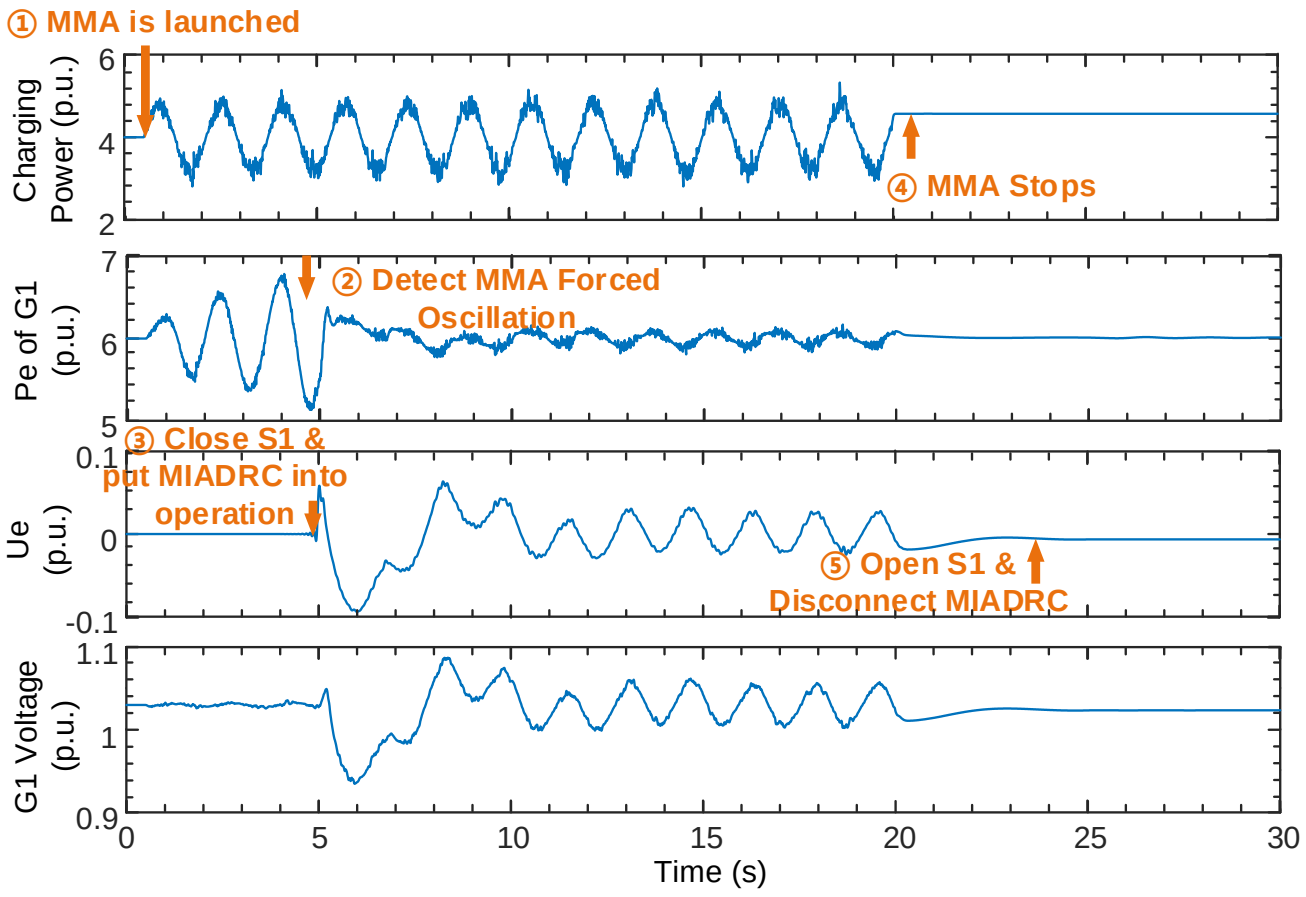

Fig. 20. The MIADRC control under the tense condition.

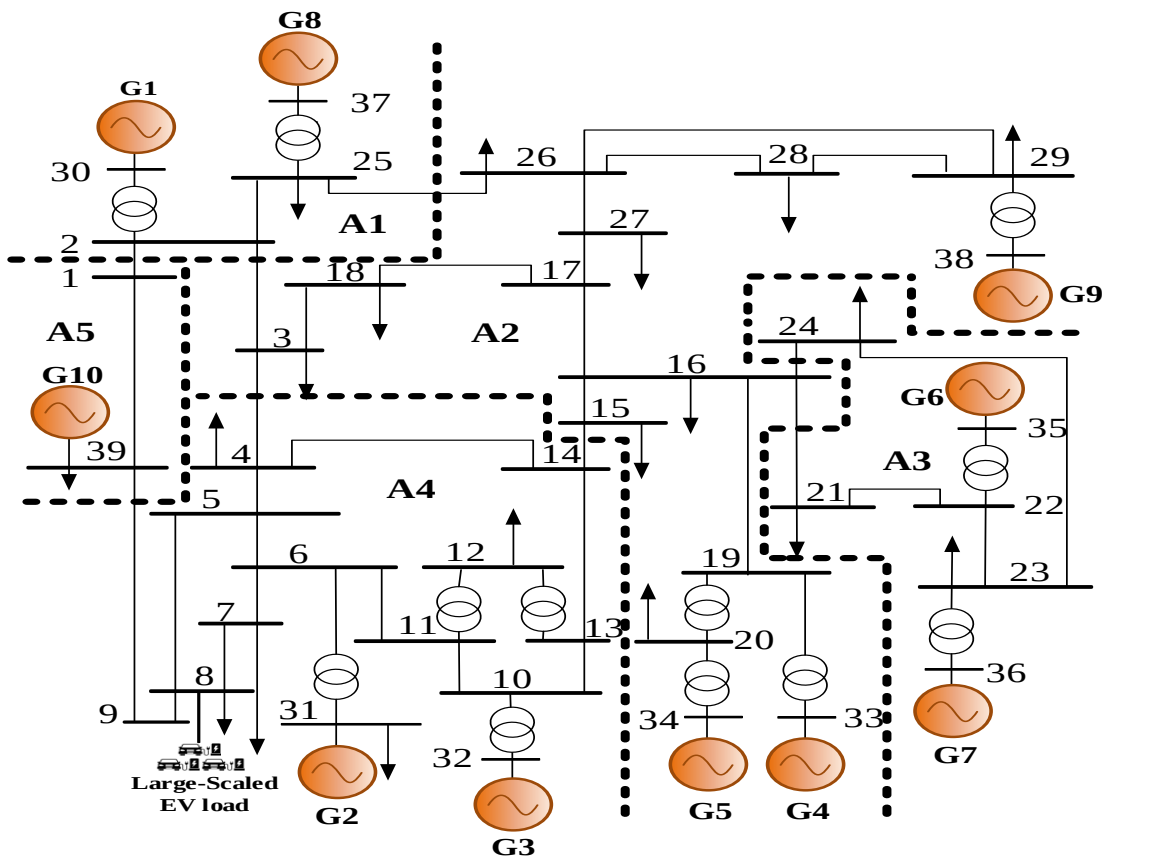

Fig. 21. The IEEE 10-machine 39-bus test system with large-scaled EV load.

At last, the stochastic effects are tested. We increased the normal random disturbance with $\Delta P_{a0}$=2.4 p.u. under the base-load scenario, and obtained the corresponding simulated (Fig. 17(a) and (b)) and calculated results (Fig. 17 (c) and (d)). As is shown, due to the weaken resonance caused by the excitation of the wideband random signal, the impact of the random disturbance is very small. Moreover, as is shown, the computed statistical variance (Fig. 17 (d)) based on the pseudo excitation method can well describe the main characteristics of the actual statistical variance (Fig. 17 (b)). Thus, the statistical variance contains the DC and the double frequency components, which is consistent with ***Law* 5**.

### *C. Performance of the MIADRC Strategy against MMA*

In this section, the performance of MIADRC against MMA is tested. Both the base-load and heavy-load conditions specified in TABLE III are considered in the 2-area 4-machine test system in Fig. 11. MMA is set to be launched at 1.0 s and sustained from 1.0 to 20.0 s. Main parameters for the MIADRC defense strategy are given below (18) and (20) with $b_i$=4600.

First, we assume that MMA is quickly detected so that MIADRC is put into operation at 1.0 s. Then the corresponding power responses generator under base and heavy load cases are simulated and shown in Fig. 18 and Fig. 19, respectively. For comparison, the defense performance of wide-area power system stabilizer (WAPSS) is also computed and given in Fig. 18 and Fig. 19, respectively. Also, the dynamic responses without defense are also demonstrated in Fig. 18 and Fig. 19, respectively. It is shown that compared with the no-defense condition, the MIADRC and WAPSS can both suppress the forced oscillations. However, the MIADRC strategy takes an obvious advantage in the suppression for the above extreme operating conditions. Hence, the MIADRC strategy has a good adaptability to different operating conditions, and MIADRC improves the system damping and actively eliminates the persistent forced oscillation generated by MMA.

Then, a time delay representing the detection time is considered to test the effectiveness of the detailed response process (Fig. 4) of the MIADRC strategy. In this test scenario, we assume the MMA detection takes 4 s, and then MIADRC can be put into operation within 0.01 s. When the forced oscillation has been suppressed, MIADRC will be disconnected automatically from the power system. The corresponding system response dynamics of this test scenario are illustrated in Fig. 20. It is shown that after MMA is detected MIADRC immediately runs after 5.0 s. And at the same time, the output signal $u_e$ of MIADRC starts increasing. As a result, the terminal voltage of G1 fluctuates in the way that compensates for the power disturbance caused by MMA. This voltage fluctuation is small but meaningful. It can be calculated from Fig. 20 that MIADRC suppresses the large forced oscillation with an amplitude of 1.008 p.u. by generating a small voltage oscillation with an amplitude of only 0.017 p.u. Compared with

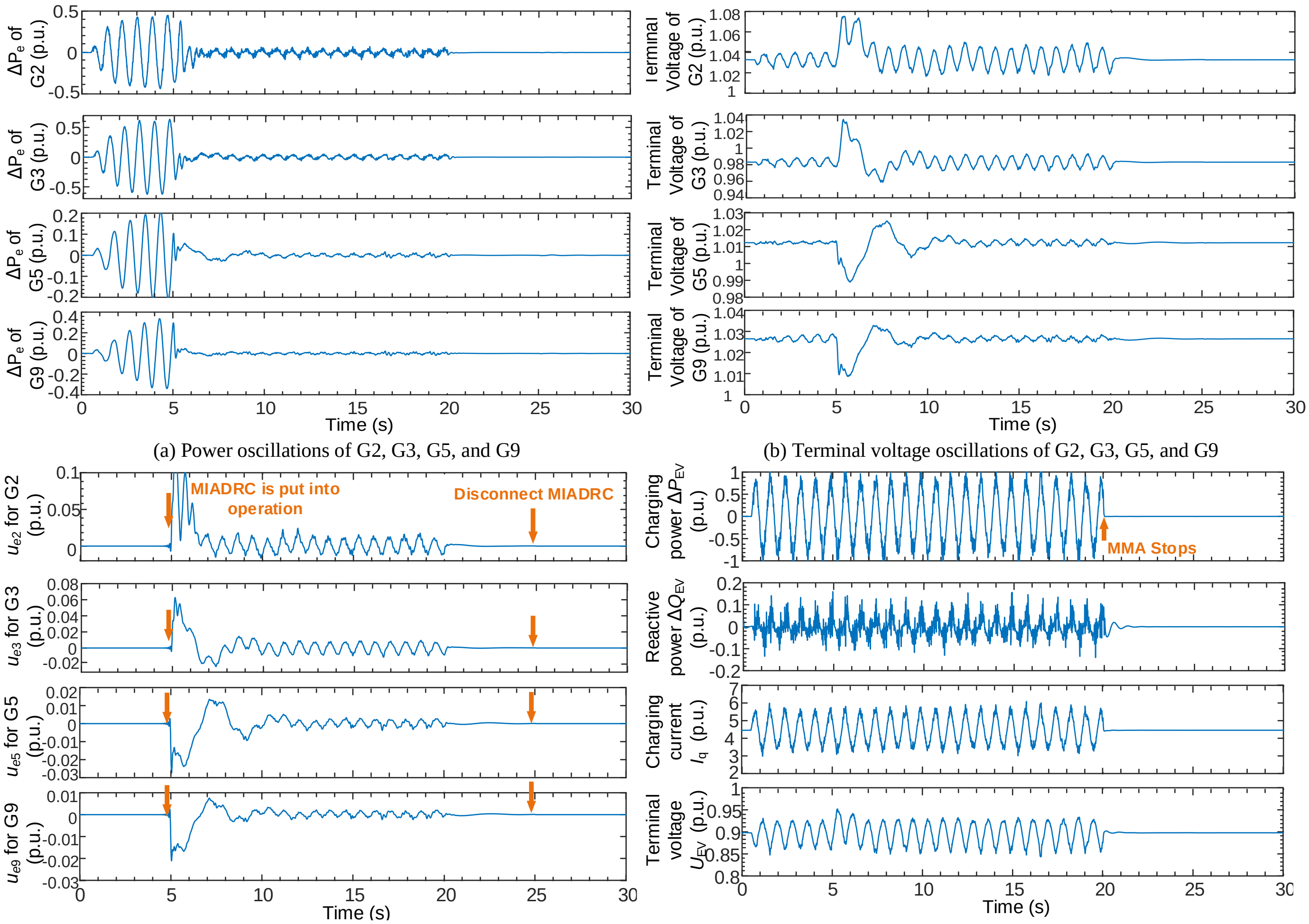

(a) Power oscillations of G2, G3, G5, and G9

(b) Terminal voltage oscillations of G2, G3, G5, and G9

(c) MIADRC control signals for G2, G3, G5, and G9

(d) Main dynamic responses of aggregative EV charging load

Fig. 21. MIADRC performance in New England 10-machine 39-bus system.

the amplitude of the forced oscillation before MIADRC is put into operation, MIADRC achieves a 91.8% suppression rate, which demonstrates the feasibility and effectiveness of the whole process of MIADRC strategy. Therefore, the proposed MIADRC strategy is effective to suppress the forced oscillation stimulated by MMA, thereby improving the power system stability.

To test the MIADRC performance in a larger test system, the IEEE New England 10-machine 39-bus system as shown in Fig. 21 is adopted. A large EV charging load is added at Bus 8. Through modal analysis, it is found that the 1.28 Hz electromechanical mode related to (G2, G3) vs. (G5, G9) is weakly damped.

We assume that MMA attacks 1.28 Hz electromechanical mode with the amplitude of 0.8 p.u., which is launched at 1.0 s and lasts until 20.0 s. The detailed response process of MIADRC strategy is also considered and adopts the same parameter as that in the 4-machine test system. The corresponding dynamics are computed and illustrated in Fig.21, where (a), (b), and (c) are power oscillations, terminal voltage oscillations, and MIADRC control signals of G2, G3, G5 and G9, respectively, and (d) gives the dynamics of the EV charging load.

It is shown that after the detection of MMA at 5.0 s, control signals from MIADRC on G2, G3, G5, and G9 immediately start generating. When MMA stops, control signals keep generating until forced oscillation is suppressed at 25.0 s. Though in a lager test system, it still shows that as soon as MIADRC is put in, obvious control signals are formed for each generator which directly reshape the terminal voltages of generators in a similar waveform. Owing to the compensation function, it shows that the corresponding output power of generators is quickly suppressed, although MMA is still attacking the charging pile to stimulate forced oscillations. Therefore, for the large power system, the performance of MIADRC in suppressing the forced oscillations caused by MMA is also validated.

## VI. Conclusion

With the controllable EV charging load developing, this paper investigated the huge threat of MMA to the power system stability, which needs to be taken seriously and investigated carefully. We mainly introduced the MMA threat, analyzed characteristics of dynamic responses to MMA, and proposed the MIADRC defense strategy. This research meets the needs of the stability of power system dynamics, and its main characteristics are concluded as below.

1) MMA can easily manipulate the charging behavior of the

charging pile through vulnerabilities to cause high-amplitude forced oscillation on the electromechanical time scale.

2) A charging pile model on the electromechanical time scale is established. And tests on the hardware platform show that the proposed model provides satisfied accuracy in the electromechanical numerical simulation and is suitable for the investigation of MMA.

3) Expressions of the mean value and standard deviation of the dynamic response to MMA are derived. And five laws of the forced oscillation stimulated by MMA are concluded. Test results verified that this forced oscillation is mainly affected by the mean response, and mainly dominated by three factors, namely, load intensity, system damping, and attack frequency.

3) Attack sources for the MMA forced oscillation is distributed in numerous charging piles and are very hard to accurately locate and remove. The proposed MIADRC defense strategy can actively feedforward the compensation and effectively suppress the forced oscillation, which improves the system damping and prevents the propagation of disturbances to the generators. Test results show that the proposed MIADRC can achieve a forced oscillation suppression rate of more than 90%.

## APPENDIX A

According to Section III.A, we have $y_i = c_{1i}\Delta u_{ti} + c_{2i}\Delta\omega_i + c_{3i}\Delta P_{ei}$, and $g_i = \left[\mathbf{0}_1^T \cdots [0 \quad 0 \quad 0 \quad k_{Ai}/T_{Ai}]_i^T \cdots \mathbf{0}_m^T\right]^T$. Then, $L_{g_j} L_f^0 y_i = \frac{\partial y_i}{\partial x^T} g_j = 0$, and $L_{g_j} L_f^1 y_i = \frac{\partial}{\partial x^T}\left(\frac{\partial y_i}{\partial x^T} f\right) g_j = c_{1i}\frac{\partial u_{ti}}{\partial E'_{qj}}\frac{1}{T'_{d0j}}\frac{K_{Aj}}{T_{Aj}} + c_{1i}\frac{\partial P_{ei}}{E'_{qj}}\frac{1}{T'_{d0j}}\frac{K_{Aj}}{T_{Aj}}$.

Considering $u_{ti} = \sqrt{(x_{qi}I_{qi})^2 + (\mathrm{E}'_{qi} - x'_{di}I_{di})^2}$, $I_{di} = -E'_{qi}B_{ii} + \sum_{j=1, j\neq i}^{n} Y_{ij}E'_{qj}\sin(\delta_{ij} - \varphi_{ij})$, $\mathrm{I}_{qi} = E'_{qi}G_{ii} + \sum_{j=1, j\neq i}^{n} Y_{ij}E'_{qj}\cos(\delta_{ij} - \varphi_{ij})$, $P_{ei} = G_{ii}{E'_{qi}}^2 + E'_{qi}\sum_{j=1. j\neq 1}^{n} E'_{qj}Y_{ij}\cos(\delta_{ij} - \varphi_{ij})$, we will get the following derivative expressions

$$\frac{\partial u_{ti}}{\partial E'_{qj}} = \begin{cases} \dfrac{{x_{qi}}^2 I_{qi}Y_{ij}\cos(\delta_{ij} - \varphi_{ij}) - (\mathrm{E}'_{qi} - x'_{di}I_{di})(1 - x'_{di}Y_{ij}\sin(\delta_{ij} - \varphi_{ij}))}{\sqrt{(x_{qi}I_{qi})^2 + (\mathrm{E}'_{qi} - x'_{di}I_{di})^2}} & i \neq j \\ \dfrac{{x_{qi}}^2 I_{qi}G_{ii} - (\mathrm{E}'_{qi} - x'_{di}I_{di})(1 + x'_{di}B_{ii})}{\sqrt{(x_{qi}I_{qi})^2 + (\mathrm{E}'_{qi} - x'_{di}I_{di})^2}} & i = j \end{cases},$$

$$\frac{\partial P_{ei}}{\partial E'_{qj}} = \begin{cases} E'_{qi}Y_{ij}\cos(\delta_{ij} - \varphi_{ij}) & i \neq j \\ 2G_{ii}E'_{qi} + \sum_{j=1. j\neq i}^{n} E'_{qj}Y_{ij}\cos(\delta_{ij} - \varphi_{ij}) & i = j \end{cases}.$$

Thus, the zero-dynamic equation of each turbine generator is

$$\begin{bmatrix} \Delta\dot{\delta}_i \\ \Delta\dot{\omega}_i \\ \Delta\dot{E}'_{qi} \\ \Delta\dot{E}_{fi} \end{bmatrix} = \begin{bmatrix} 0 & \omega_0 & 0 & 0 \\ -\dfrac{K_{1i}}{T_{Ji}} & -\dfrac{D_i}{T_{Ji}} & -\dfrac{K_{2i}}{T_{Ji}} & 0 \\ \dfrac{K_{4i}}{T'_{d0i}} & 0 & \dfrac{K_{3i}}{T'_{d0i}} & \dfrac{1}{T'_{d0i}} \\ \dfrac{K_{Ai}c_{i3}}{T_{Ai}c_{i1}}K_{1i} & \dfrac{K_{Ai}c_{i2}}{T_{Ai}c_{i1}} & \dfrac{K_{Ai}c_{i3}}{T_{Ai}c_{i1}}K_{2i} & -\dfrac{1}{T_{Ai}} \end{bmatrix} \begin{bmatrix} \Delta\delta_i \\ \Delta\omega_i \\ \Delta E'_{qi} \\ \Delta E_{fi} \end{bmatrix} \tag{A1}$$

where $K_{1i} = \frac{\partial P_{ei}}{\partial\delta_i}$, $K_{2i} = \frac{\partial P_{ei}}{\partial E'_{qi}}$, $K_{3i} = \frac{\partial E_{qi}}{\partial E'_{qi}}$, and $K_{4i} = \frac{\partial E_{qi}}{\partial\delta_i}$.

## Acknowledgments

This work was supported by the “National Natural Science Foundation of China Project” [No. 51807188], the Science Foundation of China University of Petroleum [No. 2462020YJRC008], the Hebei Province Natural Science Foundation (E2020502067) and the Fundamental Research Funds for the Central Universities (2019MS079).